\documentclass[11pt]{article}

\usepackage[margin=1in]{geometry}
\usepackage{amsmath,amssymb}
\usepackage{graphicx}
\usepackage{bm}
\usepackage{booktabs}
\usepackage[section]{placeins}
\usepackage[dvipsnames]{xcolor}
\usepackage[numbers,sort&compress]{natbib}
\usepackage{hyperref}
\hypersetup{colorlinks=true,linkcolor=RoyalBlue,citecolor=RoyalBlue,
            urlcolor=RoyalBlue}

\newcommand{\kperp}{k_\perp}

\newcommand{\dperp}{\nabla_\perp}

\title{Reduced magnetohydrodynamics with electron inertia in
pseudodipole coordinates}

\author{Khilav Majmudar, Robert L Lysak \\ \small University of Minnesota, Minneapolis, USA}
\date{}

\begin{document}
\maketitle

\begin{abstract}
\noindent
We study the altitude-dependent structuring of field-aligned currents
by driven Alfv\'en waves in an expanding auroral flux tube. The model
uses the reduced magnetohydrodynamics (RMHD) long-thin ordering assumptions 
along with electron inertia. We derive a magnetic
tension operator consistent with energy conservation in the expanding
geometry. We compare uniform density with a prescribed profile containing
a dense ionospheric end and a low-density upper region. The simulations
produce localized current concentrations near the lower boundary and
filamentary structures at greater heights. Broad structures remain
visible in the magnetic and velocity stream functions while their perpendicular
derivatives reveal finer magnetic field, velocity, and current structure. With the
nonuniform density profile, the run with additional upper-boundary
reflection shows finer current structure and greater numerical dissipation.
Spectral diagnostics show forward
nonlinear energy transfer and altitude-dependent power-law-like intervals
without a broad constant-flux range. The narrowest current concentrations
span only a few grid cells, so their physical widths remain unresolved.
\end{abstract}

\section{Introduction}

Alfv\'en waves propagating along auroral field lines carry Poynting flux from
the magnetosphere toward the ionosphere. Observations of the auroral
acceleration region show that they are accompanied by field-aligned electron
beams and small transverse scales \citep{Chaston2003,Chaston2005}.
Understanding how the associated field-aligned currents develop transverse
structure is a central problem for models of auroral electrodynamics.
The magnetic field strength, flux-tube width, and plasma density vary along
an auroral field line, changing both wave propagation and the physical
scales of the current structures. When the
perpendicular scale $k_\perp^{-1}$ approaches the electron inertial length,
electron inertia produces an appreciable parallel electric field, making the inertial Alfv\'en wave a
candidate mechanism for accelerating auroral electrons. Here we focus on
the spatial organization of the currents and fields in such a tube and
use their evolution and spectral transfer to examine the development of
small transverse scales.

The natural framework is reduced magnetohydrodynamics (RMHD) \citep{Strauss1976,1974JETP...38..283K}
which was first introduced to study the stability of plasmas in large aspect-ratio tokamaks.
Here the ordering assumes a strong guide field and perpendicular scales much shorter
than parallel ones. Under this so called long, thin approximation, the equations of MHD get simplified into a system of
two nonlinear PDEs in the velocity and magnetic stream functions. 
The assumptions separating various MHD systems (compressible vs incompressible, high vs low plasma $\beta$)
are discussed systematically by \citet{zank1992equations}.
This is a suitable setting for flux tubes at auroral latitudes in the inner magnetosphere. Applying it to
a tube whose cross-section expands with altitude introduces a subtlety
absent from the straight-cylinder case. It is useful to label each background
field line by where it intersects the lower boundary. These fixed transverse
labels describe a grid that widens with the tube.
Nonuniform-tube RMHD formulations such as those used in studies of the solar corona already account for this changing geometry
\citep{vanBallegooijen2011}. Here we apply such a geometry to an
electron-inertial auroral model. \citet{Seyler1990} included electron inertial effects in
a reduced MHD model of small-scale discrete auroral arcs. \citet{Chaston2010} used the same model to
study auroral current sheet structuring under the influence of instabilities.
These auroral calculations use a uniform background field and rectangular
coordinates. \citet{Chaston2010} relate their transverse sizes to other
altitudes using dipolar scale factors, while retaining a uniform field in
the dynamical equations. This is distinct from evolving perturbations in
a continuously expanding tube with altitude-dependent transverse derivatives.

We examine the resulting current structure at different heights under
continuous Alfv\'enic driving. Uniform density runs provide a reference
for a profile with a dense ionospheric end and decreasing density above
it. Variations in numerical diffusion and upper boundary reflection
allow us to examine the sensitivity of the current concentrations and
their finer structure. Transverse and longitudinal sections of the
magnetic and velocity potentials, fields, and currents connect the
large-scale driving pattern to the smaller-scale spatial variations.

The expanding geometry requires consistent magnetic tension and induction
terms. Accounting for the changing cross-section in the energy balance
gives the tension term $B_0^2\partial_s(\mathcal J/B_0)$, which reduces
to the familiar $B_0\partial_s\mathcal J$ in \citet{Seyler1990} when
$B_0$ is constant.

Wave reflection and boundary escape determine how much of the injected
energy remains available for nonlinear evolution. Gradients in Alfv\'en
speed can reflect waves even though the local dispersion relation has
no propagation cutoff. The ionospheric boundary also reflects waves
\citep{Lysak1985,lysak1991feedback}. We distinguish these mechanisms in
linear calculations before examining the driven nonlinear evolution.

Spectra and nonlinear energy-transfer rates complement the spatial
current maps by measuring how energy is distributed and transferred
across transverse scales. We distinguish approximately power-law
spectra from a constant-flux cascade range, and examine the competing
effects of parallel transport and numerical dissipation. Neither a
spectral slope nor a narrow current feature alone establishes a
resolved turbulent cascade or a particular current-sheet instability.

We first specify the geometry, reduced equations, and energy balance,
then present the numerical method and linear reflection results. The
nonlinear energy budgets and spectra establish the context for the
current-structure analysis, which is the main focus of this work.
We compare current and field morphology across height, density profile
and boundary conditions, and conclude with the implications and
limitations of these current-structure results.

\section{Model}

\subsection{Geometry}
We work in a paraxial flux tube whose cross-sectional area expands with
altitude at fixed magnetic flux. Here paraxial means that the transverse
size $L_\perp$ is small compared with the background variation length
defined as follows.
\begin{equation}
H_B=\frac{B_0}{|d B_0/ds|}=\left | \frac{d}{ds} \ln \left ( \frac{B_0}{B_I} \right ) \right |^{-1};
\qquad
\frac{L_\perp}{H_B}\ll1
\end{equation}
We retain the leading order 
flaring out of the background field but neglect curvature of the tube axis and
higher order geometric corrections as in thin-tube treatments of nonuniform
RMHD of \citet{vanBallegooijen2011}. This captures dipolar area expansion but
not the full curvature of a dipole field. Such a formalism is adequate to 
describe flux tubes at polar latitudes and at low altitudes up to around 2-3 $R_E$. 
The perturbation ordering is
specified separately in Section~\ref{sec:ordering}.
Writing $s$ for the
arc length along the background field and $z$ for altitude above the
ionosphere, we get at retained order the longitudinal coordinate as
$s=z$ and the physical transverse coordinates as $x=a(s)\xi$ and
$y=a(s)\eta$. The scale factor $a(s)$ makes the quantities 
$x/a(s)$, $y/a(s)$ remain constant along a field line. 
Demanding that the magnetic field follows an inverse cubed profile
and that the flux is conserved over all altitudes gives the following relations:
\begin{equation}
  B_0(s)=B_I\left(\frac{R_I}{R_I+s}\right)^{3}
  \qquad
  a(s)=\left(\frac{B_I}{B_0(s)}\right)^{1/2}
      =\left(\frac{R_I+s}{R_I}\right)^{3/2}
  \label{eq:B0}
\end{equation}
The corresponding paraxial background in physical Cartesian coordinates is
\begin{equation}
 \mathbf B_{\rm bg}=B_0(s)\mathbf e_z
 -\frac{B_0'(s)}{2}(x\mathbf e_x+y\mathbf e_y).
 \label{eq:background-vector}
\end{equation}
The transverse components provide the tube expansion which is set by the divergence-free condition
$\nabla\cdot\mathbf B_{\rm bg}=B_0'+2B_0a'/a=0$.
The coordinates $(\xi,\eta)$ label the ionospheric footpoint of a background field
line since $a(0)=1$. Differentiation along the tube is performed at
fixed $(\xi,\eta)$ while the same label interval corresponds to a wider
physical interval as $a(s)$ increases. This explicitly carries the
cross-section dependence emphasized in nonuniform-tube RMHD
\citep{vanBallegooijen2011} into our coordinates.

The perpendicular gradient and Laplacian are respectively
\begin{equation}
 \nabla_\perp=a^{-1}(\mathbf e_\xi\partial_\xi+\mathbf e_\eta\partial_\eta),
 \qquad \nabla_\perp^2=a^{-2}(\partial_\xi^2+\partial_\eta^2).
 \label{eq:perp-metric}
\end{equation}
For periodic label widths $L_\xi,L_\eta$, a transverse Fourier mode is
$\exp[i(\kappa_\xi\xi+\kappa_\eta\eta)]$, where
$\kappa_\xi=2\pi m/L_\xi$ and $\kappa_\eta=2\pi n/L_\eta$ for integers
$m,n$. Its physical wavenumbers are
\begin{equation}
 k_x(s)=\frac{2\pi m}{a(s)L_\xi},\quad
 k_y(s)=\frac{2\pi n}{a(s)L_\eta},\quad
 k_\perp^2(s)=\frac{4 \pi^2}{a^2(s)}
 \left(\frac{m^2}{L_\xi^2}+\frac{n^2}{L_\eta^2}\right).
 \label{eq:physical-wavenumber}
\end{equation}
Thus a fixed pair of mode numbers
has $k_\perp^2\propto a^{-2}=B_0/B_I$. For the \(128^2\) simulation grid, 
the physical transverse spacing is \(\Delta x(s)=\Delta y(s)=a(s)\Delta x_I\), 
with \(\Delta x_I=0.25\) km at the ionospheric boundary. At \(s=9900\) km, 
\(a=4.0902\), giving \(\Delta x=\Delta y=1.0225\) km. The circular two-thirds 
spectral truncation retains \(k_\perp<2\pi/[3\Delta x(s)]\); the largest actual 
retained Fourier modes therefore have \(k_{\perp,\max}=8.372\,\mathrm{km}^{-1}\) 
at the bottom and \(2.047\,\mathrm{km}^{-1}\) at the top.
The volume element is
$dV=J_g\,d\xi\,d\eta\,ds$ with the Jacobian as
\begin{equation}
 J_g=\det\!\left[\frac{\partial(x,y,z)}{\partial(\xi,\eta,s)}\right]
     =a^2(s)=\frac{B_I}{B_0(s)}.
 \label{eq:jacobian}
\end{equation}
This is the physical area represented by one unit of flux-label area.

\begin{figure}[htbp]
\centering
\includegraphics[width=.8\textwidth]{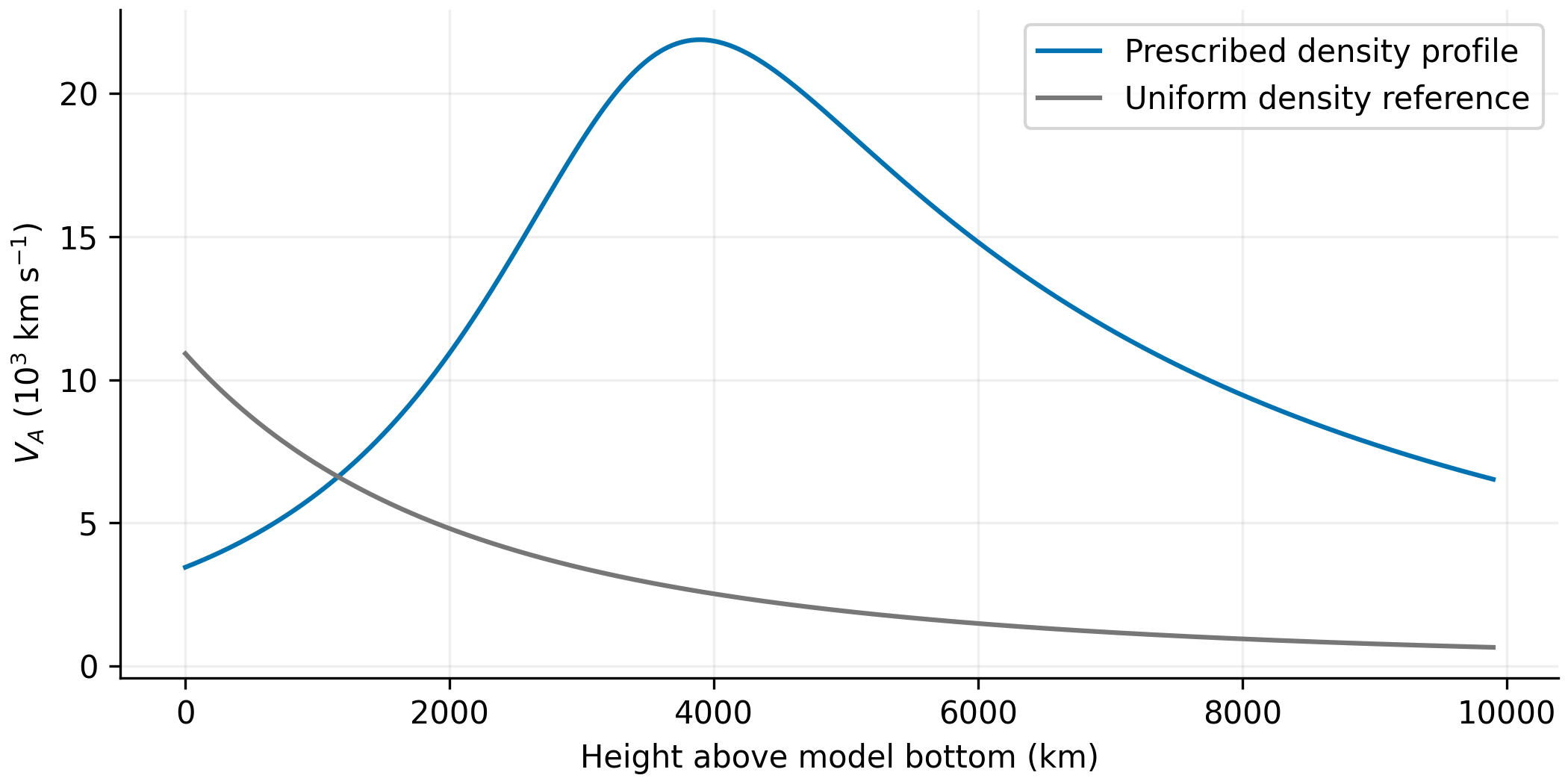}
\caption{Alfv\'en speed in the same expanding magnetic geometry for the
uniform density reference and the non-uniform density profile of
Eq.~\eqref{eq:density-profile}. The density enhancement lowers the speed
at the ionospheric end and produces an interior maximum.}
\label{fig:density-alfven}
\end{figure}

In the reference calculations we take $n_e=10^4\,\mathrm{cm^{-3}}$ to be uniform.
The Alfv\'en speed then varies
only through $B_0$, falling from $10\,915\ \mathrm{km\,s^{-1}}$ at the
ionospheric footpoint to $652\ \mathrm{km\,s^{-1}}$ at $9900$~km altitude.
Uniform density deliberately excludes the ionospheric density enhancement.
In a more realistic profile, the dense ionosphere lowers $V_A$ and the
subsequent rise of $V_A$ above it partially reflects waves. Together with reflection
at the lower conducting boundary, this can form an ionospheric Alfv\'en
resonator \citep{lysak1991feedback}.
For example, \citet{Chaston2010} use two density components to represent the
variation of $V_A$. To include this effect here, we
use a static vertically-varying proton--electron background with an 
exponential profile given by 
\begin{equation}
 n_e(s)=\left[100+99900\exp\!\left(-\frac{s}{500\,\mathrm{km}}\right)\right]
       \,\mathrm{cm^{-3}},\qquad \rho(s)=m_p n_e(s).
 \label{eq:density-profile}
\end{equation}
Here $s=z$ is
height above the model lower boundary. 
Figure~\ref{fig:density-alfven} compares the resulting Alfv\'en speeds.
This density profile gives $V_A=3451\,\mathrm{km\,s^{-1}}$ at the bottom,
a maximum of $21884\,\mathrm{km\,s^{-1}}$ near $3900$~km, and
$6524\,\mathrm{km\,s^{-1}}$ at the top. Its long-wavelength one-way
transit time is $\int_0^{9900\,\mathrm{km}}ds/V_A=0.991$~s.

\subsection{Derivation}
\label{sec:ordering}
We use the quasineutral two-fluid dynamics with electron
inertia retained in Ohm's law\citep{Seyler1990}. With singly charged ions, $n_i=n_e=n$, bulk velocity
$\mathbf v\simeq\mathbf v_i$, and $\rho\simeq m_i n$, the nondissipative
parent equations can be written in SI units as
\begin{align}
 \partial_t n+\nabla\cdot(n\mathbf v)&=0,
 \label{eq:parent-continuity}\\
 \rho(\partial_t+\mathbf v\cdot\nabla)\mathbf v
   &=-\nabla(p_i+p_e)+\mathbf j\times\mathbf B,
 \label{eq:parent-momentum}\\
 \partial_t\mathbf B&=-\nabla\times\mathbf E,\qquad
 \mu_0\mathbf j=\nabla\times\mathbf B,\qquad\nabla\cdot\mathbf B=0,
 \label{eq:parent-maxwell}\\
 \mathbf E+\mathbf v_e\times\mathbf B
   &=-\frac{\nabla p_e}{en}
     -\frac{m_e}{e}(\partial_t+\mathbf v_e\cdot\nabla)\mathbf v_e,
 \qquad\mathbf v_e=\mathbf v-\frac{\mathbf j}{en}.
 \label{eq:parent-ohm}
\end{align}
Here $e$ is the elementary charge and $p_i,p_e$ are scalar species pressures. Ion mass contributes to the
bulk inertia and electron inertia is kept where it permits a parallel electric
field. These parent equations describe more degrees of
freedom than are analyzed in the RMHD calculations. For example, we do not describe
compressive dynamics like sound waves and magnetosonic waves.

We select the low-frequency shear-Alfv\'en sector using the ordering
\begin{equation}
 \frac{\ell_\perp}{\ell_\parallel}\sim
 \frac{|\mathbf b_\perp|}{B_0}\sim
 \frac{|\mathbf v_\perp|}{V_A}=O(\epsilon)\ll1,
 \qquad \frac{\omega}{\Omega_i}\ll1,
 \qquad \Omega_i=\frac{eB_0}{m_i}.
 \label{eq:rmhd-ordering}
\end{equation}
The low-frequency ordering \(\omega/\Omega_i\ll1\) 
is well satisfied by the imposed driving. The minimum proton gyrofrequency occurs at 
the tube top: \(\Omega_{i,\min}=eB_{\min}/m_p=286.3\,\mathrm{s}^{-1}\), or \(f_{ci,\min}=45.56\) Hz. 
Thus the production driving frequency \(f=0.1\) Hz gives \(\omega/\Omega_i=f/f_{ci}\le2.20\times10^{-3}\) 
throughout the tube, independently of temperature.

The parallel direction includes background and wave variation.
Bulk parallel flow, density perturbations and parallel magnetic
perturbations are taken to be of higher order in $\epsilon$. As mentioned above, fast compressive waves and
their pressure/density evolution are not retained
\citep{zank1992equations,vanBallegooijen2011}.
We also assume cold electrons and an ion gyroradius small compared to 
the retained perpendicular scales. This applies 
to the inertial Alfv\'en wave limit. 

For a cold proton background with \(k_B T_{i\perp}\le0.35\) eV
motivated by the cold polar-wind temperatures measured by \citet{drakou1997ion}, 
the thermal gyroradius \(\rho_i=\sqrt{m_pk_BT_{i\perp}}/(eB_0)\) 
remains small compared with the retained transverse scales. 
At the tube top, \(B_0=2.989\,\mu\mathrm T\) gives \(\rho_i\le20.2\) m. 
With \(k_{\perp,\max}=2.047\,\mathrm{km}^{-1}\), the maximum \(k_\perp\rho_i\le0.0414\), 
supporting dropping of ion gyroradius effects.

We omit electron pressure and Hall coupling. 
Electron inertia is retained by allowing $k_\perp\lambda_e$ up to order
unity in the equations, with
\begin{equation}
 \lambda_e=\frac{c}{\omega_{pe}}
   =\sqrt{\frac{m_e}{\mu_0 n_e e^2}},\qquad
 \omega_{pe}=\sqrt{\frac{n_e e^2}{\epsilon_0m_e}}.
 \label{eq:electron-length}
\end{equation}
For the $128^2$ grids, the circular spectral cutoff gives a maximum
$k_{\rm cut}\lambda_e\simeq0.445$ in the uniform-density model and
$1.88$ with the non-uniform density profile. Thus the latter includes
perpendicular scales comparable to the local electron inertial length.

At leading retained order the transverse flow is divergence-free. Taking
the field aligned component of the curl of Eq.~\eqref{eq:parent-momentum}
eliminates the scalar pressure gradient.
The perpendicular part of Ohm's law supplies the leading $\mathbf E\times
\mathbf B$ motion while its parallel electron inertia term and Faraday's
law lead to the induction equation. Keeping the transverse advection and
magnetic tension brackets gives the two-field electron-inertial system
\citep{Seyler1990}. The geometric factors for the nonuniform background
are constrained by its area-weighted energy exchange derived below.

\subsection{Evolved equations and numerical diffusion}
Define
\begin{equation}
  \mathbf{v}_\perp=\mathbf{e}_s\times\dperp U,
  \qquad
  \mathbf{b}_\perp=\mathbf{e}_s\times\dperp A,
  \qquad
  \Omega=\dperp^{2}U,
  \qquad
  \mathcal{J}=\dperp^{2}A,
\end{equation}
and $[f,g]=\mathbf{e}_s\cdot(\dperp f\times\dperp g)$, the reduced system with
electron inertia is
\begin{align}
  \partial_t\Omega+[U,\Omega]
    &=\frac{1}{\mu_0\rho}\left\{[A,\mathcal{J}]
      +B_0^{2}\,\partial_s\!\left(\frac{\mathcal{J}}{B_0}\right)\right\},
    \label{eq:vorticity}\\[4pt]
  \left(\partial_t+[U,\,\cdot\,]\right)\!
    \left(A-\lambda_e^{2}\mathcal{J}\right)
    &=\partial_s(B_0U),
    \label{eq:induction}
\end{align}
with $\lambda_e$ the electron inertial length and $\rho$ the proton mass
density. The non-uniform density is still unvarying in the perpendicular 
directions and so the commutator brackets do not generate additional 
terms with density gradients. 

These are the nondissipative reduced evolution equations.
In the transverse direction we use a Fourier pseudospectral method, where derivatives
are calculated in Fourier space and nonlinear products in physical space
(Section~\ref{sec:numerics}). The nonlinear runs also
use fourth-order perpendicular numerical hyperdiffusion,
$\mathcal D_\nu U$ and $\mathcal D_\eta A$ in the evolution of $U$ and $A$. 
They are numerical hyperviscosity and magnetic hyperdiffusivity with 
Fourier multipliers $-\nu_2 k_\perp^4$ and $-\eta_2 k_\perp^4$ respectively. 
Magnetic diffusion acts after the inversion of the $1-\lambda_e^2\nabla_\perp^2$ operator in the induction equation~\eqref{eq:induction}.

\subsection{The tension operator}
\label{sec:tension}
Equation~\eqref{eq:vorticity} has the tension term as
$B_0^{2}\partial_s(\mathcal{J}/B_0)$.

The operator is derived by requiring that the two
equations exchange energy consistently. Take the time derivative of the energy~\eqref{eq:energy}
and integrate by parts in the perpendicular plane. Using the physical
volume element and its Jacobian $J_g=a^2$ from Eq.~\eqref{eq:jacobian},
we get the kinetic and magnetic parts as  
$-\int J_g\,\rho(s)U\,\partial_t\Omega$ and $-\mu_0^{-1}\int J_g\,\mathcal{J}\,\partial_t
(A-\lambda_e^2\mathcal J)$ respectively.
Thus with a trial tension
operator $\mathcal{T}[\mathcal{J}]$ in Eq.~\eqref{eq:vorticity} and
$\partial_s(B_0U)$ in Eq.~\eqref{eq:induction} we obtain
\begin{equation}
  \left.\frac{dE}{dt}\right|_\parallel
  =-\frac{1}{\mu_0}\int\! J_g
    \Bigl[\,U\,\mathcal{T}[\mathcal{J}]
    +\mathcal{J}\,\partial_s(B_0U)\Bigr]d\xi\,d\eta\,ds .
  \label{eq:exchange}
\end{equation}
For this expression to integrate to an end flux, the 
integrand must be an $s$-derivative. Use $J_g=B_I/B_0$ with the constant
ionosphere field $B_I$ and the quantities
\begin{equation}
  X\equiv B_0U=\Phi,\qquad Y\equiv\frac{\mathcal{J}}{B_0}.
  \label{eq:conj}
\end{equation}
Then the choice $\mathcal{T}[\mathcal{J}]=B_0^{2}\partial_s(\mathcal{J}/B_0)$ gives
\begin{equation}
  \frac{B_I}{B_0}\Bigl[U B_0^{2}\partial_s Y+\mathcal{J}\,\partial_sX\Bigr]
  =B_I\Bigl[X\,\partial_sY+Y\,\partial_sX\Bigr]
  =B_I\,\partial_s\!\left(XY\right)
  \label{eq:perfect}
\end{equation}
Thus Eq.~\eqref{eq:exchange} reduces to the boundary flux in
Eq.~\eqref{eq:poynting} below.

The pairing in Eq.~\eqref{eq:conj} has a geometric interpretation.
For a small cross-sectional area $dS$, the field aligned current is
$dI=j_\parallel dS$ and the background magnetic flux is
$d\mathcal F=B_0dS$. Consequently
\begin{equation}
 \frac{dI}{d\mathcal F}=\frac{j_\parallel}{B_0}
   =\frac{\mathcal J}{\mu_0 B_0}=\frac{Y}{\mu_0}.
 \label{eq:current-per-flux}
\end{equation}
Thus $\mathcal J/B_0$ is proportional to current per unit background
magnetic flux. It weights current by the area of a fixed-flux element as that element
expands. 

Now the quadratic energy
\begin{equation}
  E=\frac{1}{2}\int J_g\,d\xi\,d\eta\,ds
    \left[\rho\left|\dperp U\right|^{2}
    +\frac{\left|\dperp A\right|^{2}
      +\lambda_e^{2}\left|\dperp^{2}A\right|^{2}}{\mu_0}\right]
  \label{eq:energy}
\end{equation}
obeys
\begin{equation}
  \frac{dE}{dt}=-\bigl[P\bigr]_{s_1}^{s_2},
  \qquad
  P=\frac{J_gB_0}{\mu_0}\int d\xi\,d\eta\;U\mathcal{J},
  \label{eq:poynting}
\end{equation}
so the energy changes only by Poynting flux through the ends.

Writing $q(s)=k_\perp^2(s)$ for a single transverse Fourier mode,
$q=(\kappa_\xi^2+\kappa_\eta^2)B_0/B_I$ by
Eq.~\eqref{eq:physical-wavenumber}. Thus the correct operator reduces to
\begin{equation}
  B_0^{2}\,\partial_s\!\left(-\frac{qA}{B_0}\right)=-B_0\,q\,\partial_sA,
\end{equation}
which contains no term proportional to $A/(R_I+z)$.

The corresponding term in the induction equation is unaffected. If we expand
\begin{equation}
  \partial_s(B_0U)=B_0\left(\partial_sU-\frac{3U}{R_I+z}\right)
\end{equation}
we get a $-3U/r$ contribution.

\subsection{Linear dispersion relation}
The local dispersion relation obtained by linearising Eqs.~\eqref{eq:vorticity}--\eqref{eq:induction}
holding the background coefficients at their local values is
\begin{equation}
  \omega^{2}=\frac{k_\parallel^{2}V_A^{2}}{1+\kperp^{2}\lambda_e^{2}},
  \qquad
  V_A=\frac{B_0}{\sqrt{\mu_0\rho}} .
  \label{eq:disp}
\end{equation}
This local dispersion relation contains no
propagation cutoff. 
For the uniform density linear tests, the continuum comparison below identifies partial reflection from the
background gradient, measured in Section~\ref{sec:reflection}. A measure of
the background variation over the parallel phase scale is
\begin{equation}
  \epsilon_B\equiv\frac{1}{|k_\parallel|H_B}
  =\frac{3}{|k_\parallel|(R_I+z)}
  =\frac{3V_A}{|\omega|(R_I+z)\sqrt{1+k_\perp^2\lambda_e^2}} .
  \label{eq:epsB}
\end{equation}
For $\epsilon_B\ll1$ the background magnetic field
changes little over the phase scale $|k_\parallel|^{-1}$, supporting a local
WKB description. $\epsilon_B\gtrsim1$ indicates that this scale
separation fails and gradient reflection may be important. For our model,
at the driving frequency \(f=0.1\) Hz and for 
\(k_\perp\lambda_e\ll1\), \(\epsilon_B\) ranges from \(0.19\) to \(8.19\) 
in the uniform density model and from \(1.91\) to \(10.31\) in the nonuniform model.
In this long-wavelength limit, $\epsilon_B$ normalizes only the magnetic-gradient
term in $k_\parallel'/k_\parallel$ and does not include the density-gradient
term. We emphasize that failure of WKB scale separation does not by itself violate RMHD
ordering which requires parallel background and perturbation scales to
remain long compared with perpendicular scales, not with each other.

\section{Numerical method}
\label{sec:numerics}

The transverse directions are treated using the pseudospectral method with $2/3$
dealiasing~\citep{Orszag1971} using FFTW~\citep{FFTW2005}. For the square
$N_x\times N_x$ grids, the circular truncation retains integer Fourier
mode pairs $(m,n)$ satisfying $\sqrt{m^2+n^2}<N_x/3$.
This excludes modes that could introduce aliasing errors from the
quadratic nonlinear products into the retained modes.
The perpendicular derivatives 
are calculated in Fourier space by multiplication with $ik_x$ and $ik_y$. Transverse multiplications are 
performed in physical space and the product is transformed back to 
Fourier space before continuing with the calculations. The parallel
direction uses a staggered grid with $U$ on nodes $1\ldots n_z$ and $A$ on
faces $1\ldots n_z-1$. Time integration uses the fourth-order Runge--Kutta method.
The fixed timestep is based on the shortest parallel Alfvén crossing time 
between grid points via the CFL condition. Perpendicular advection and explicit hyperdiffusion are 
also monitored for timestep restrictions.

\section{Linear results}

\subsection{Gradient reflection}
\label{sec:reflection}
We drive the top boundary with a single Fourier mode
($\kperp\lambda_e=0.0104$) at fixed flux label. The lower boundary is kept
transparent and the reflection characteristics at the
driven end are measured. The result is compared with a separate 
frequency-domain solution of the linear equations integrated as a two-point
boundary value problem.

\begin{table}[htbp]
\centering
\caption{Simulated and reference power reflectances as a function of driving
frequency $f$. $\epsilon_{\rm flux}$ is the normalized power balance error and
$N_\lambda$ is the number of grid spacings per parallel wavelength at the upper boundary.}
\label{tab:reflection}
\begin{tabular}{ccccc}
\toprule
$f$ (Hz) & $R_{\rm sim}$ & $R_{\rm ref}$ & $\epsilon_{\rm flux}$ & $N_\lambda$ \\
\midrule
0.01 & 0.78042 & 0.78042 & $1.2\times10^{-6}$ & 652 \\
0.1  & 0.38970 & 0.38857 & $4.8\times10^{-4}$ & 65 \\
0.5  & 0.08596 & 0.07781 & $2.8\times10^{-3}$ & 13 \\
1.0  & 0.04752 & 0.02930 & $1.3\times10^{-2}$ & 6.5 \\
\bottomrule
\end{tabular}
\end{table}
The power reflectance and transmittance are defined by
\begin{equation}
 R=\frac{\langle P_{\rm reflected}\rangle}
         {\langle P_{\rm incident}\rangle},
 \qquad
 T=\frac{\langle P_{\rm transmitted}\rangle}
         {\langle P_{\rm incident}\rangle}.
 \label{eq:power_reflection_transmission}
\end{equation}
Here the magnitudes of the powers are integrated over the tube
cross-section and averages are taken over complete driving
periods. Incident and reflected powers are evaluated at the upper boundary
and transmitted power at the transparent lower boundary.
$R_{\rm sim}$ is measured in the time-dependent simulation and 
$R_{\rm ref}$ is obtained from the frequency-domain linear boundary-value
problem. The normalized power-balance error is
$\epsilon_{\rm flux}=|R_{\rm sim}+T_{\rm sim}-1|$.
The parallel resolution is measured by
$N_\lambda=\lambda_{\parallel,\rm top}/\Delta z
\simeq V_{A,\rm top}/(f\Delta z)$, where the last expression neglects the
electron inertia correction. The upper boundary has
the shortest local parallel wavelength and thus the fewest grid spacings
per wavelength. 

Table~\ref{tab:reflection} shows reflectance falling monotonically.
The high-frequency cases show disagreement with the reference linear cases
due to their limited parallel grid resolution.

\subsection{Ionospheric boundary conditions}
\label{sec:ionosphere}
The lower boundary imposes the condition $U=Z_{\rm bc}A$ with
$Z_{\rm bc}=(\mu_0\Sigma_PB_I)^{-1}$ for a height-integrated
Pedersen conductance $\Sigma_P$~\citep{Lysak1985}. The reflection 
coefficient is given by 
\begin{equation}
  r=\frac{Z_{\rm bc}-Z_A}{Z_{\rm bc}+Z_A},
  \label{eq:refl}
\end{equation}
\[
Z_A=\frac{V_{A,I}}{B_I}
      \sqrt{1+k_{\perp,I}^{\,2}\lambda_{e,I}^{\,2}}.
\] 
$Z_A$ is the local Alfv\'en characteristic coefficient. $|r|\le1$ for nonnegative
$\Sigma_P$. 

For each transverse mode, impedance matching requires $r=0$, giving
\begin{equation}
 \Sigma_{P,\mathrm{match}}(k_{\perp,I})
 =\frac{1}{\mu_0 V_{A,I}
          \sqrt{1+k_{\perp,I}^{\,2}\lambda_{e,I}^{\,2}}}.
 \label{eq:matched-conductance}
\end{equation}
In the long-wavelength limit this reduces to the Alfv\'en conductances
$\Sigma_A=1/(\mu_0V_{A,I})$, which is $0.0729$~S for the uniform density
reference and $0.231$~S for the non-uniform density profile.
Conductances above or below the mode's matching value produce reflection.
For the uniform density reference in this limit, $\Sigma_P=1$--$10$~S
gives power reflectances $|r|^2=0.75$--$0.97$.

\section{Nonlinear transfer in the driven tube}

\subsection{The configuration}
The uniform density production results have a $128\times128\times100$ grid
with a $32$~km square cross-section at the ionosphere and a
$9900$~km radial extent. This extent follows from the $100$ parallel grid
points separated by $\Delta s=100$~km, so the $99$ intervals give
$L_\parallel=(N_z-1)\Delta s=9900$~km. Table~\ref{tab:production} lists the
dimensional inputs. The top incoming wave is driven continuously
throughout the $120$~s run at $f=0.1$~Hz. The intervals studied here are
therefore during driven evolution and not freely decaying turbulence. The spatial
driver in footpoint flux coordinates is the bipolar profile
\begin{equation}
 g(\xi,\eta)=\phi_0\left[
 e^{-[(\xi-d)^2+\eta^2]/w^2}
 -e^{-[(\xi+d)^2+\eta^2]/w^2}\right],
 \quad w=4~\mathrm{km},\quad d=6~\mathrm{km},
 \label{eq:driver}
\end{equation}
with sinusoidal time dependence and the same spectral truncation as the
fields. Here $d$ is the offset of each lobe and $\phi_0$ is the stream-function forcing amplitude.
At the top, widths and offsets are larger by $a(9900)=4.0902$.
The bottom is fully reflecting and the top imposes no additional reflected
incoming characteristic.

\begin{table}[htbp]
\centering
\caption{Expanding tube simulation parameters. Numerical diffusion is
varied at fixed resolution and forcing. $\eta_2=\nu_2$ in every case.}
\label{tab:production}
\begin{tabular}{ll}
\toprule
Quantity & Value\\
\midrule
$R_I$, $B_I$ & $6356$~km, $50\,000$~nT\\
$n_e$, $\lambda_e$ & $10^4, \mathrm{cm^{-3}}$ $\sim$ $0.0531$~km\\
$\Delta x_I=\Delta y_I$, $\Delta s$ & $0.25$~km, $100$~km\\
$\phi_0$, $f$, duration & $600~\mathrm{km^2\,s^{-1}}$, $0.1$~Hz, $120$~s\\
Diffusion order $p$ & $2$\\
$\nu_2$ ($\mathrm{km^4\,s^{-1}}$) & $0.00529$, $0.0132$, $0.0279$, $0.0588$\\
Reference-case $\nu_2$ & $0.0279~\mathrm{km^4\,s^{-1}}$\\
\bottomrule
\end{tabular}
\end{table}

\subsection{Spectral flux}
The perpendicular energy spectrum $E(\kperp)$ describes how fluctuation
energy is distributed among transverse spatial scales.
A forward energy cascade is a dynamical process in which nonlinear
interactions transfer energy from larger to smaller scales. Its diagnostic
is the spectral energy flux which is the net rate at which nonlinear interactions
transfer energy across a chosen wavenumber from modes below it to modes
above it. This transfer through scale space is distinct from the transport
of wave energy along the magnetic field \citep{BrunoCarbone2013}.

In the classical, statistically stationary cascade picture, energy enters
at large scales (injection scales) and is removed by dissipation at small scales. 
The range of scales between the injection and dissipation scales is called the inertial range. 
Here, direct forcing and
dissipation are negligible and there is little mean energy accumulation.
Approximately the same energy per unit time passes through each
intermediate scale, giving a nonzero and nearly constant mean spectral flux
\citep{AlexakisBiferale2018}. 

Thus, a power law spectrum alone does not establish a cascade. It
computes the energy content but not the direction or rate of transfer. For example,
equilibrium spectra can have power law forms with zero mean spectral flux
\citep{AlexakisBiferale2018}. In our open tube, parallel transport
can also supply or remove energy at a given height and scale. We therefore
compare nonlinear transfer with parallel transport and dissipation, 
distinguishing power-law-like spectral shapes from intervals
consistent with an inertial range cascade.

We define $T(\kperp)$ as the rate of change of energy in a perpendicular
wavenumber shell due to the nonlinear terms of
Eqs.~\eqref{eq:vorticity}--\eqref{eq:induction}. After moving the nonlinear
terms to the right-hand side, we retain the 
advection, Lorentz force and induction terms. The perpendicular
Laplacian and electron-inertia inversions contribute to
$\partial_t U$ and $\partial_t A$. Nonlinear products are evaluated
pseudospectrally and transformed back to Fourier space. There, each
nonlinear rate is multiplied by the complex conjugate of its corresponding
field coefficient and the mode weight from the quadratic energy. Taking
the real part and summing over modes in a shell gives its nonlinear
energy-transfer rate. As positive $T$ denotes an energy gain by that shell, 
we define the cumulative transfer with a minus
sign so that a net nonlinear loss from modes below $\kperp$ corresponds to
positive transfer toward smaller transverse scales:
\begin{equation}
  \Pi(\kperp)=-\!\!\sum_{\kperp'\le\kperp}\!\!T(\kperp') .
  \label{eq:flux}
\end{equation}
Summed over the tube the budget is
\begin{equation}
  \frac{\partial E(\kperp)}{\partial t}
    = T(\kperp)+F(\kperp)+D(\kperp),
  \label{eq:budget}
\end{equation}
where $F$ is the net energy input through the tube ends from the linear
parallel coupling and $D\le0$ is the energy removed by numerical
perpendicular dissipation. At an individual height the parallel coupling
describes energy arriving from or leaving for neighbouring layers.
Summing over height thus leaves only the end fluxes.
The nonlinear terms conserve energy within each transverse plane 
in the continuum equations. However, when discretized, the staggered placement of velocity and magnetic 
variables introduces contributions that cancel between neighbouring layers and leave 
a boundary remainder. 
Thus $\Pi$ cannot automatically be interpreted as
energy transfer solely between perpendicular scales. We assess candidate
cascade intervals together with the parallel transport, dissipation and
energy storage, rather than from the spectrum or $\Pi$ alone.

Over $100$--$120$~s, the reference tube receives a mean net boundary
power of $314.36$~kW through the linear parallel coupling, while numerical
perpendicular dissipation removes $339.66$~kW. Including the discrete
nonlinear boundary remainder of $-6.47$~kW gives a predicted decrease
in stored energy of $31.77$~kW, compared with the measured decrease of
$32.92$~kW. The difference is $1.15$~kW or $0.34\%$ of the dissipated
power. Thus dissipation is supplied by both the net incoming power and
a decrease in stored energy. The power integrals are estimated from
snapshots separated by $0.5$~s; this agreement is subject to temporal
sampling error and does not establish a statistically stationary state.

\subsection{Competition between escape and transfer}
\label{sec:escape}
Our simulations comprise eight $128^2\times100$ runs with continuous
$0.1$~Hz forcing and two driving amplitudes,
$\phi_0=450$ and $600\ \mathrm{km^2\,s^{-1}}$. At each amplitude we use
four choices of the numerical hyperdiffusion coefficients,
$\nu_2=\eta_2=0.00529$, $0.0132$, $0.0279$ and
$0.0588~\mathrm{km^4\,s^{-1}}$ with hyperdiffusion order $p=2$.
The lower boundary imposes $U=0$ and the driven upper boundary 
does not have additional reflections. 

To distinguish incoming and outgoing wave power, we separate the
Alfv\'en characteristics $U+Z_AA$ and $U-Z_AA$. Here the 
electron inertial characteristic coefficient is
\begin{equation}
  Z_A(s,k_\perp)=\frac{V_A(s)}{B_0(s)}
       \sqrt{1+k_\perp^2(s)\lambda_e^2},
  \label{eq:local-impedance}
\end{equation}
which is the same one used at the ionosphere in
Section~\ref{sec:ionosphere}.
At the top, $U+Z_AA$ is incoming and $U-Z_AA$ is outgoing and 
the difference of their directional powers gives us the net linear boundary
input. At the bottom boundary, $U=0$ makes the incident and reflected
powers equal, so no wave power is absorbed there.
The boundary fields give a gross top input of
$3.2006$~MW and outgoing power $2.881$--$2.910$~MW in the late
$\phi_0=600$ runs. $90.0$--$90.9\%$ escapes through the top, leaving
about $0.29$--$0.32$~MW net input. Over $100$--$120$~s, perpendicular numerical dissipation is
$0.267$--$0.340$~MW.
Most incoming wave power is balanced by outgoing power through the upper 
boundary leaving a much smaller net input to the tube. This escape may 
constrain the nonlinear transfer but the energy budget alone does not 
determine why the simulations lack a broad inertial range. Numerical 
diffusion and finite spatial resolution may also contribute to this effect.

The analysis below uses the complete energy-transfer records from the four
$\phi_0=600$ runs. 

\begin{figure}[htbp]
\centering
\includegraphics[width=\textwidth]{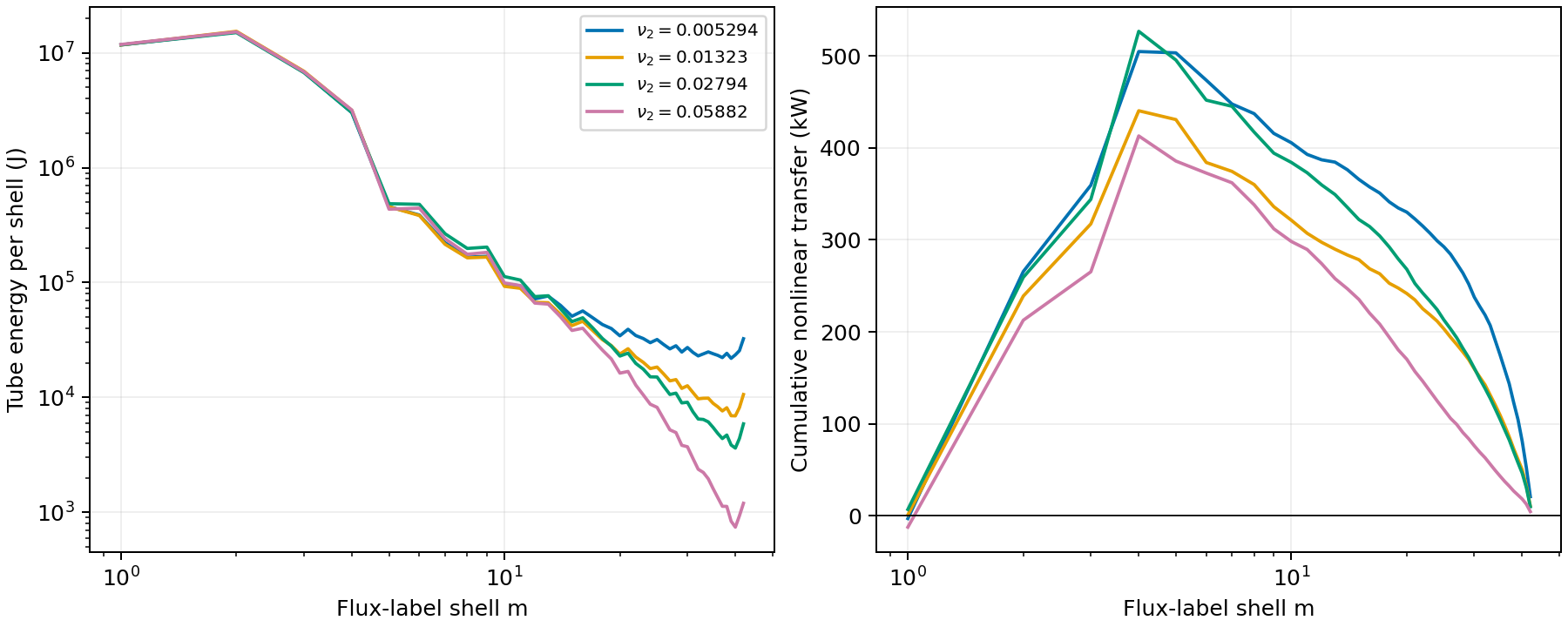}
\caption{Tube-integrated shell energy and cumulative nonlinear transfer 
averaged over $100$--$120$~s at $128^2$.
Positive cumulative transfer indicates net removal of energy from shells
below the plotted index by the nonlinear terms. A constant-flux cascade
would additionally require an approximately horizontal interval in $\Pi$
with the same power passing through successive perpendicular scales. Such 
a broad plateau is not present here. Internal redistribution would
give zero cumulative transfer after summing over all modes. Here the
nonzero endpoint includes a discrete nonlinear boundary contribution and
the omission of 40 retained outer Fourier modes from the saved shell
diagnostic. These modes are evolved in the simulation but absent from
this plot. The endpoint is not artificially set to zero and must not be
interpreted as power cascading beyond the resolved scales.
The coefficients in the legend have units $\mathrm{km^4\,s^{-1}}$, and
$\eta_2=\nu_2$ in each run.}
\label{fig:latest}
\end{figure}

\section{Spectral shape and transfer}
\label{sec:spectra}
We distinguish a descriptive power-law-like spectrum from a resolved local
inertial range. A different exponent at a different altitude is physically
allowed as the tube changes structure and hence length scales with height.

\begin{figure}[htbp]
\centering
\includegraphics[width=0.95\textwidth]{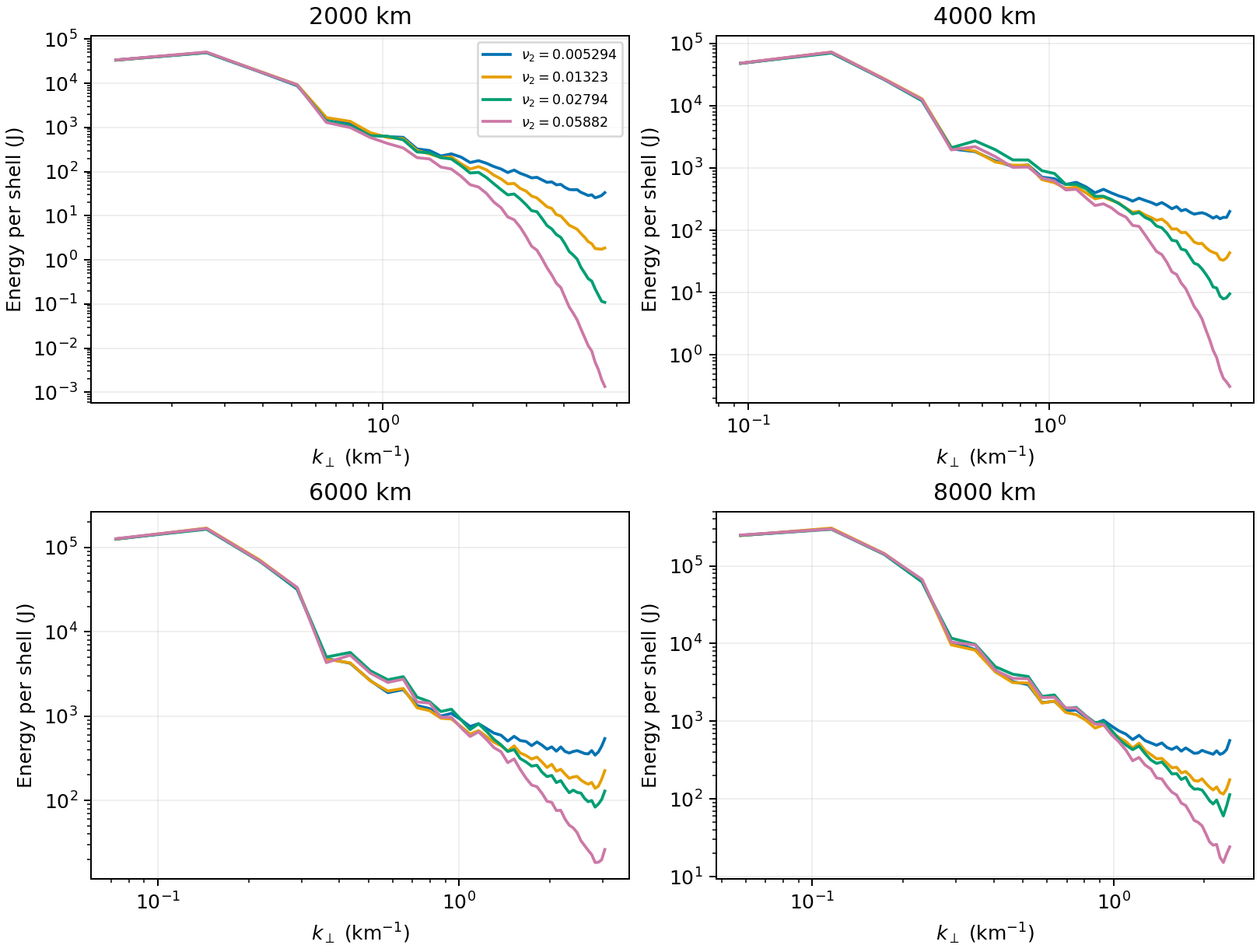}
\caption{Perpendicular energy spectra at four altitudes on the $128^2$
grid, averaged over $100$--$120$~s. Each point gives the energy in one
wavenumber shell within a tube segment of longitudinal thickness
$\Delta s=100$~km centred at the stated altitude.
Shell index $m$ corresponds to
$k_\perp=2\pi m/[32a(z)]$ in $\mathrm{km^{-1}}$.
All stored shells are shown. Slope fits are restricted to $m\le34$,
leaving a chosen $20\%$ buffer below the dealiasing cutoff $N_x/3$. 
Numerical diffusion affects the small-scale spectrum differently
at different altitudes.}
\label{fig:spectra}
\end{figure}

We identify approximately power-law spectral intervals whose slopes persist
between two averaging windows using the selection criteria in
Appendix~\ref{app:band-selection}. Table~\ref{tab:bands} lists representative intervals.
The span in decades is the base-ten logarithm of the wavenumber ratio,
$\log_{10}(k_{\perp,\max}/k_{\perp,\min})$, so
half a decade corresponds to a factor of $\sqrt{10}\simeq3.16$.
The widest band (fitted wavenumber interval) contains 29 shells, $m=6$--$34$, at $6000$~km
for $\nu_2=\eta_2=0.02794~\mathrm{km^4\,s^{-1}}$.
Its shell-centre wavenumbers extend from $k_{\perp,\min}\simeq0.435$
to $k_{\perp,\max}\simeq2.463~\mathrm{km^{-1}}$, giving
$\log_{10}(34/6)=0.753$ decades. These descriptive intervals are not by
themselves evidence of an inertial range. To interpret them as a cascade
one also requires the nonlinear transfer and energy budget.

\begin{table}[htbp]
\centering
\caption{Representative persistent descriptive bands with $\eta_2=\nu_2$.
These are values of the fitted indices.}
\label{tab:bands}
\begin{tabular}{cccccc}
\toprule
$\nu_2$ & $z$ (km) & shells & decades & $80$--$100$ s & $100$--$120$ s\\
\midrule
0.01323 & 5000 & 6--25 & 0.620 & $-1.873$ & $-1.685$\\
0.01323 & 6000 & 5--27 & 0.732 & $-1.768$ & $-1.754$\\
0.01323 & 7000 & 5--26 & 0.716 & $-1.821$ & $-1.995$\\
0.02794 & 6000 & 6--34 & 0.753 & $-2.349$ & $-2.211$\\
0.02794 & 7000 & 6--31 & 0.713 & $-2.376$ & $-2.307$\\
0.02794 & 8000 & 6--23 & 0.584 & $-2.471$ & $-2.394$\\
0.05882 & 7000 & 6--22 & 0.564 & $-2.491$ & $-2.365$\\
\bottomrule
\end{tabular}
\end{table}

\begin{figure}[htbp]
\centering
\includegraphics[width=\textwidth]{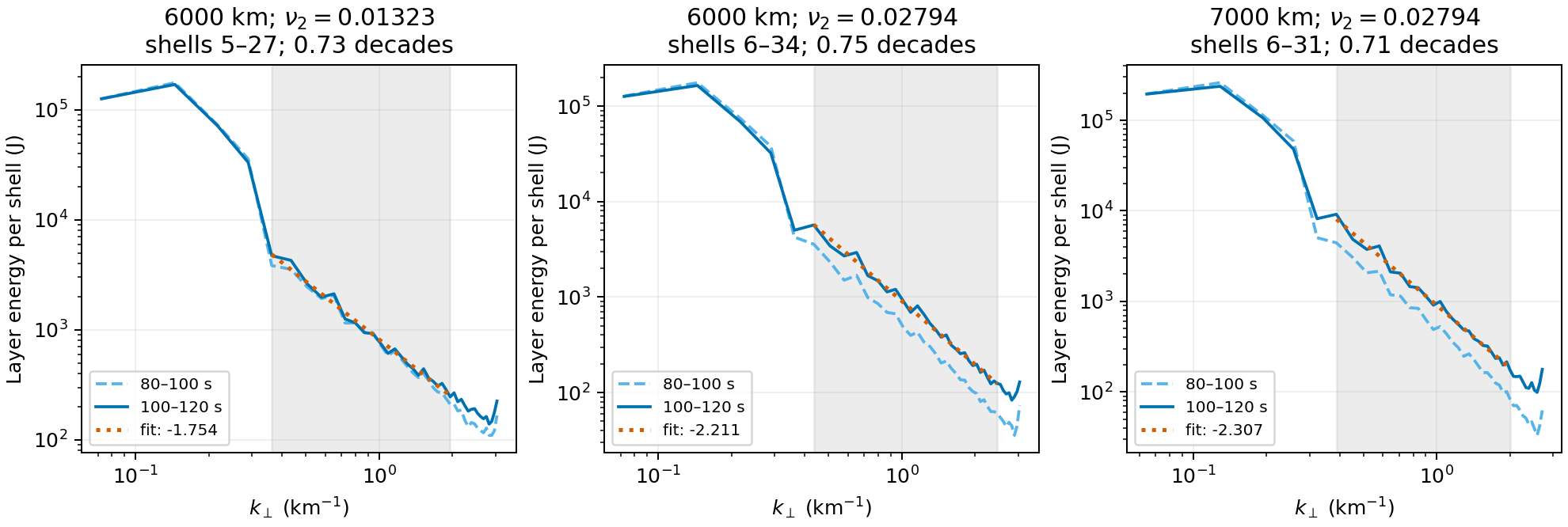}
\caption{Illustrative power-law-like bands from Table~\ref{tab:bands}.
Shading marks the tested interval and the dotted yellow fit is over the $100$--$120$~s window.
}
\label{fig:bands}
\end{figure}

Figure~\ref{fig:bands} shows three illustrative bands with their full tails
and both averaging windows. The overall spectral amplitude can change between 
the two averaging intervals even when the fitted slope remains similar.

Each $20$~s averaging interval covers only two periods of the imposed
driving. Similar slopes in the two intervals show that the spectral shape
persists over this time, but do not establish a statistical steady state
or precise uncertainties in the slopes. We therefore do not use small
differences between fitted slopes to distinguish between turbulence models.

Our retained Fourier modes are found by rounding their radial index
$\sqrt{m^2+n^2}$ to the nearest integer. But we store only shells from 
$1$ to $\lfloor N_x/3\rfloor$. At $128^2$, 40 modes satisfy the dealiasing
condition $\sqrt{m^2+n^2}<128/3\simeq42.67$ yet round to shell 43.

\subsection{Uniform slab comparison}
For comparison, we use a slab with a uniform $130.886$~km square cross-section.
This is equal to the widening tube's top width. The driver width and offset scale by $4.0902$
with the box. With $p=2$, $\nu_2=\eta_2=7.8194$ preserves the original damping rate
$\nu_2k_{\rm cut}^4$. Density and driving amplitude are unchanged and 
ionospheric field of $50\,000$~nT is kept as the constant background. Compare this to 
the pseudodipole tube where at the top boundary the background field is $2989$~nT.
At fixed density and $\kperp$, the impedance
$Z_A$ in Eq.~\eqref{eq:local-impedance} is independent of $B_0$. 
The incoming wave power thus scales
linearly with $B_0$ as per Eq.~\eqref{eq:poynting}. The ratio of its
average over one driving period in the uniform slab and in the widening tube
is $50\,000/2989\simeq16.73$.
This comparison matches the upper boundary width but 
not its magnetic field or incident power and so the spectral differences 
cannot be attributed solely to the expanding geometry.

The wide slab is evolved to $120$~s using two different timesteps. 
The total energy,
dissipation and shell spectra agree to within $4.4\times10^{-9}$ relative to each other.

The uniform slab shows a sharply falling spectrum. During
$100$--$120$~s, $<0.01\%$ of its stored energy lies in shells 7 and
higher. This is essentially unchanged from an earlier time window of $10$--$30$~s. 
The maximum cumulative nonlinear transfer across any shell is 
$0.162$~MW.
For comparison the $\nu_2=0.02794$ pseudodipole tube has $\sim 4\%$ of stored energy in
these shells and maximum cumulative transfer $0.527$~MW in the same window.
The nondispersive one-way Alfv\'en transit times are 
$0.91$~s in the wide slab and $6.08$~s in the tube, so a $120$~s simulation run corresponds
to about 132 and 20 such transits. Electron inertia also
makes the propagation time mode dependent.

\begin{figure}[htbp]
\centering
\includegraphics[width=\textwidth]{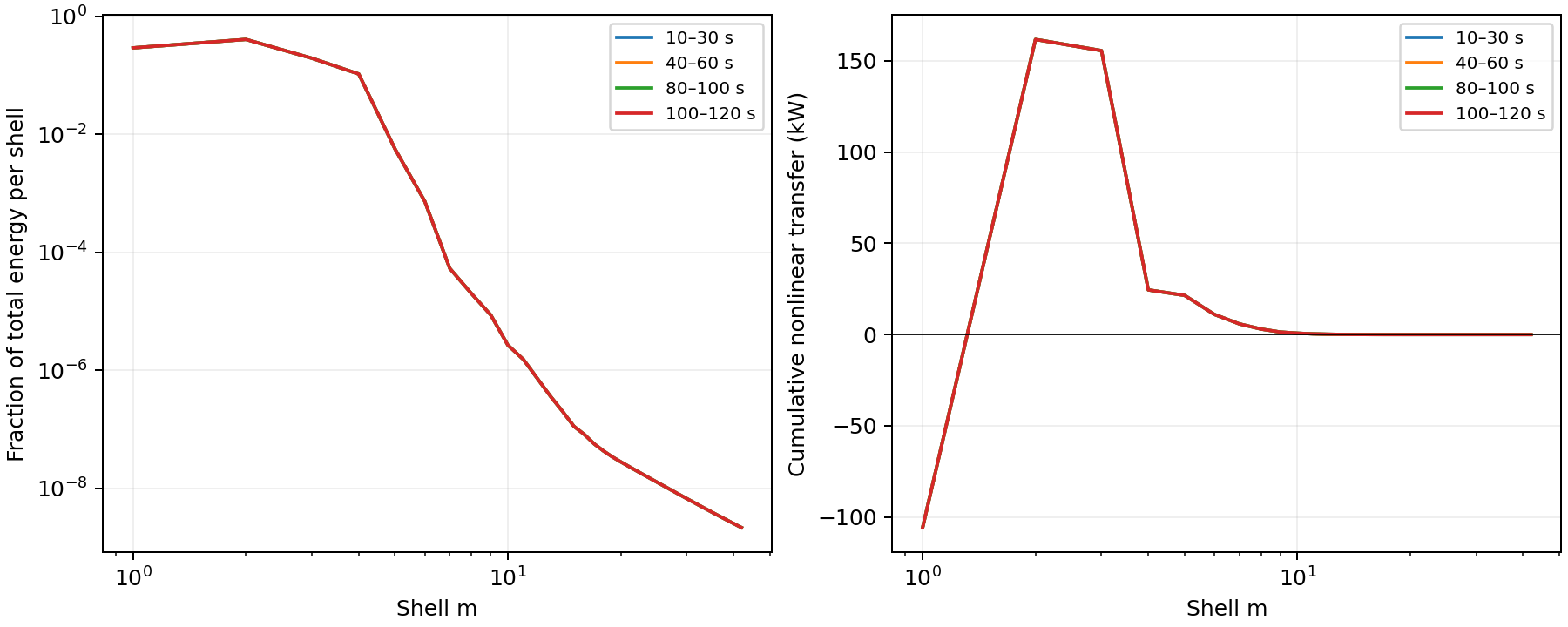}
\caption{Shell energy and cumulative nonlinear transfer for the uniform slab
at four time windows.
Extending the run to $120$~s did not populate a broad
high-shell range.}
\label{fig:wide}
\end{figure}

Figure~\ref{fig:wide} shows the uniform slab's shell-energy distribution 
and cumulative nonlinear transfer over four averaging intervals 
between 10 and 120~s. The curves nearly overlap, indicating little 
subsequent evolution of these quantities. The spectrum remains concentrated 
at low shell indices.

\subsection{Density profile and upper boundary reflection}
\label{sec:density-transfer}
We next use the density profile in Eq.~\eqref{eq:density-profile} with
the same $128^2$ transverse grid, $100$-km parallel spacing and spatial
driver shape as the reference tube. Its amplitude is rescaled to
$\phi_0=1899.22\,\mathrm{km^2\,s^{-1}}$ to match with 
the reference's mean incident
power, evaluated using the retained driver modes and the new local
characteristic impedance at the top. The open-top run lasts $120$~s and the run with
upper boundary reflection lasts $40$~s. Both are continuously driven at
$0.1$~Hz with a perfectly conducting lower boundary and 
$\nu_2=\eta_2=0.027938\,\mathrm{km^4\,s^{-1}}$.
For the reflecting top, the incoming wave satisfies
$C_{\rm down}=d+rC_{\rm up}$, where $d$ is the driver and
$r=0.5$. This leads to a $25\%$ power reflection.

The longer open run retains the height dependent power-law-like intervals in the energy spectrum.
Shells $5$--$34$ satisfy the spectral shape criteria in both
$80$--$100$ and $100$--$120$~s at $3000$, $4000$ and $5000$~km.
This interval spans $\log_{10}(34/5)=0.833$ decades and at $4000$~km its two
fitted slopes are $-2.31$ and $-2.28$. At $6000$~km the common interval
is shells $5$--$26$, with slopes $-2.12$ and $-2.03$.
Reflection broadens the fitted interval at some heights---for example
shells $4$--$34$ at $3000$~km, or $0.929$ decades. 
These descriptive fits do not establish a universal spectral index.

\begin{figure}[htbp]
\centering
\includegraphics[width=\textwidth]{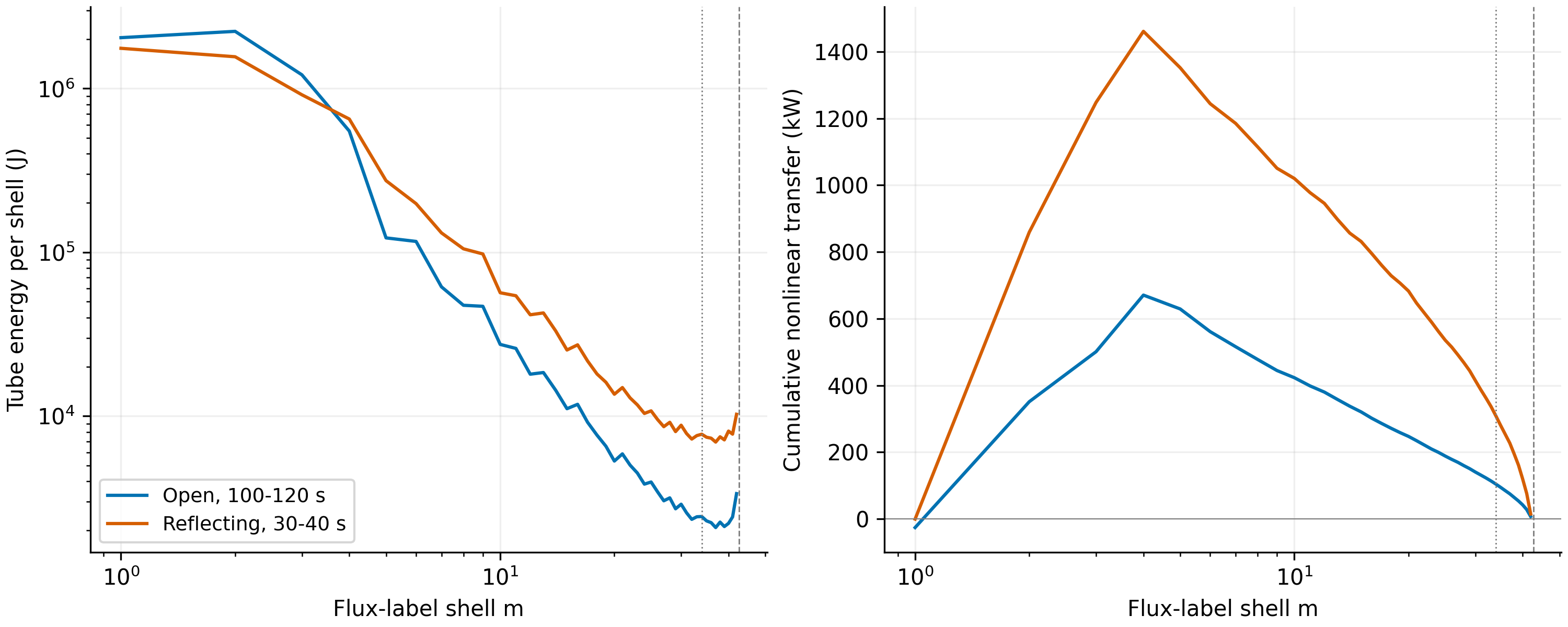}
\caption{Tube-integrated shell energy and cumulative nonlinear transfer
for the density profile runs in the format of Fig.~\ref{fig:latest}.
The open run is averaged over $100$--$120$~s and the reflecting run over
$30$--$40$~s.
Positive transfer denotes net nonlinear removal from the lower
shells in the sum. A broad horizontal interval would indicate approximately
constant flux through successive scales.
Dotted lines mark the fitting margin $m=34$ and dashed lines the circular
cutoff $N_x/3$.}
\label{fig:density-transfer}
\end{figure}

Figure~\ref{fig:density-transfer} shows tube-integrated spectra and transfer. 
Reflection increases high-wavenumber
energy and nonlinear transfer but neither run develops a broad constant-flux
plateau. The upturn near the last stored shells indicates sensitivity to
the truncation region.
Note that dotted fitting margin at $m=34$ is not the same as the circular dealiasing
cutoff at radial mode number $\sqrt{m_x^2+m_y^2}=N_x/3=42.67$, 
whose physical wavenumber is $k_{\rm cut}(s)=2\pi/[3\Delta x(s)]$.

We also compare the
same $30$--$40$-s interval of these two runs. Mean stored energy is
$6.582$~MJ for the open top and $6.243$~MJ for the reflecting top whereas
mean numerical dissipation rises from $317$ to $742$~kW. Net linear
boundary input rises from $278$ to $726$~kW.
At $100$--$120$~s the open run has $6.656$~MJ stored energy and
$256$~kW numerical loss. Based on the data from these averaging windows, we can 
say that reflection increases the rate at which
energy reaches dissipative scales without increasing the retained mean energy.

\section{Current structure and auroral interpretation}
\label{sec:current}
The driven tube develops current concentrations whose strength and
transverse shape vary with altitude. We examine their relation to the
magnetic and velocity structures, first in the uniform-density reference
and then with the dense ionospheric profile. Earlier auroral RMHD studies
have examined current structuring and its relation to instabilities
\citep{Seyler1990,Chaston2010}. Here the emphasis is on the spatial
structure produced within the expanding tube.

For our conventions the field-aligned current is
\begin{equation}
 j_\parallel=\mathcal{J}/\mu_0=\nabla_\perp^2 A/\mu_0.
\end{equation}

\subsection{Altitude dependence in the uniform density tube}
\begin{figure}[htbp]
\centering
\includegraphics[width=\textwidth]{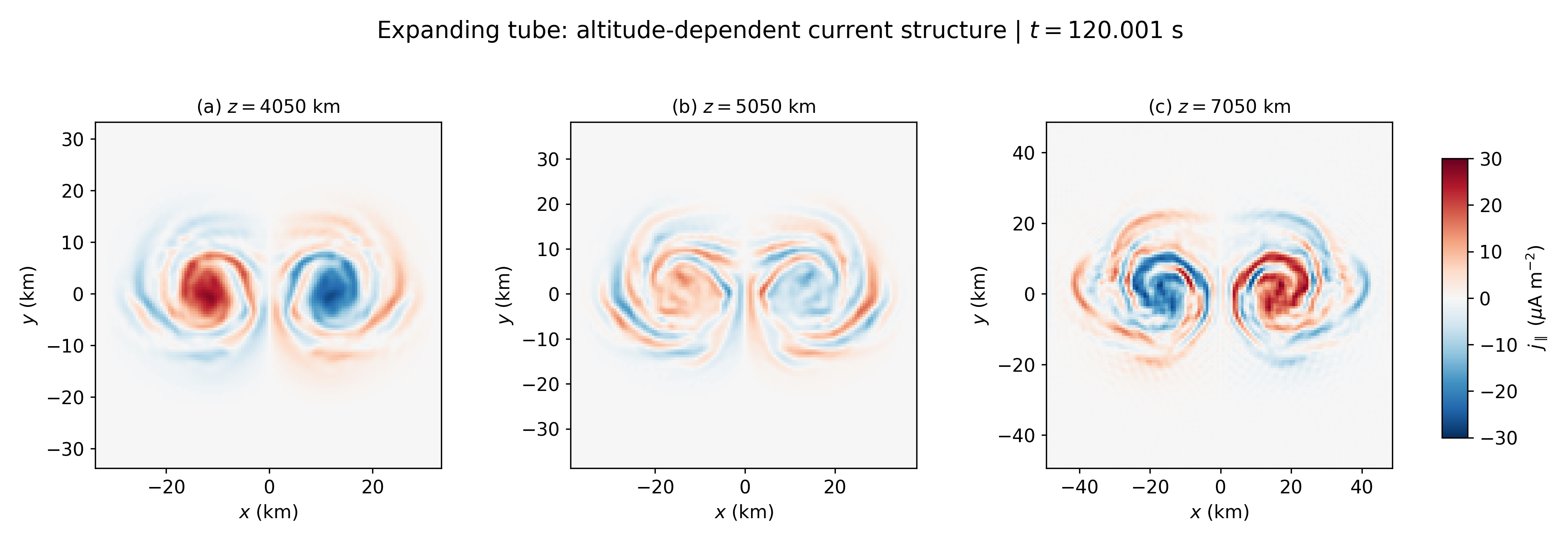}
\caption{Signed field-aligned current in the uniform-density reference tube at
$t=120$~s and $z=4050$, $5050$ and $7050$~km. Each panel spans the
full local cross-section.}
\label{fig:current-altitudes}
\end{figure}

The uniform density simulations show localized currents near the
footpoint and curved, filamentary structures at greater heights.
For the reference run, $\nu_2=\eta_2=0.02794~\mathrm{km^4\,s^{-1}}$,
Fig.~\ref{fig:current-altitudes} shows transverse sections at three
altitudes. Bands of positive and negative field-aligned current surround
the main concentrations, with shapes and relative strengths that change
along the tube. Each panel spans the local cross-section, so its physical
width increases with height.

\begin{figure}[htbp]
\centering
\includegraphics[width=0.95\textwidth]{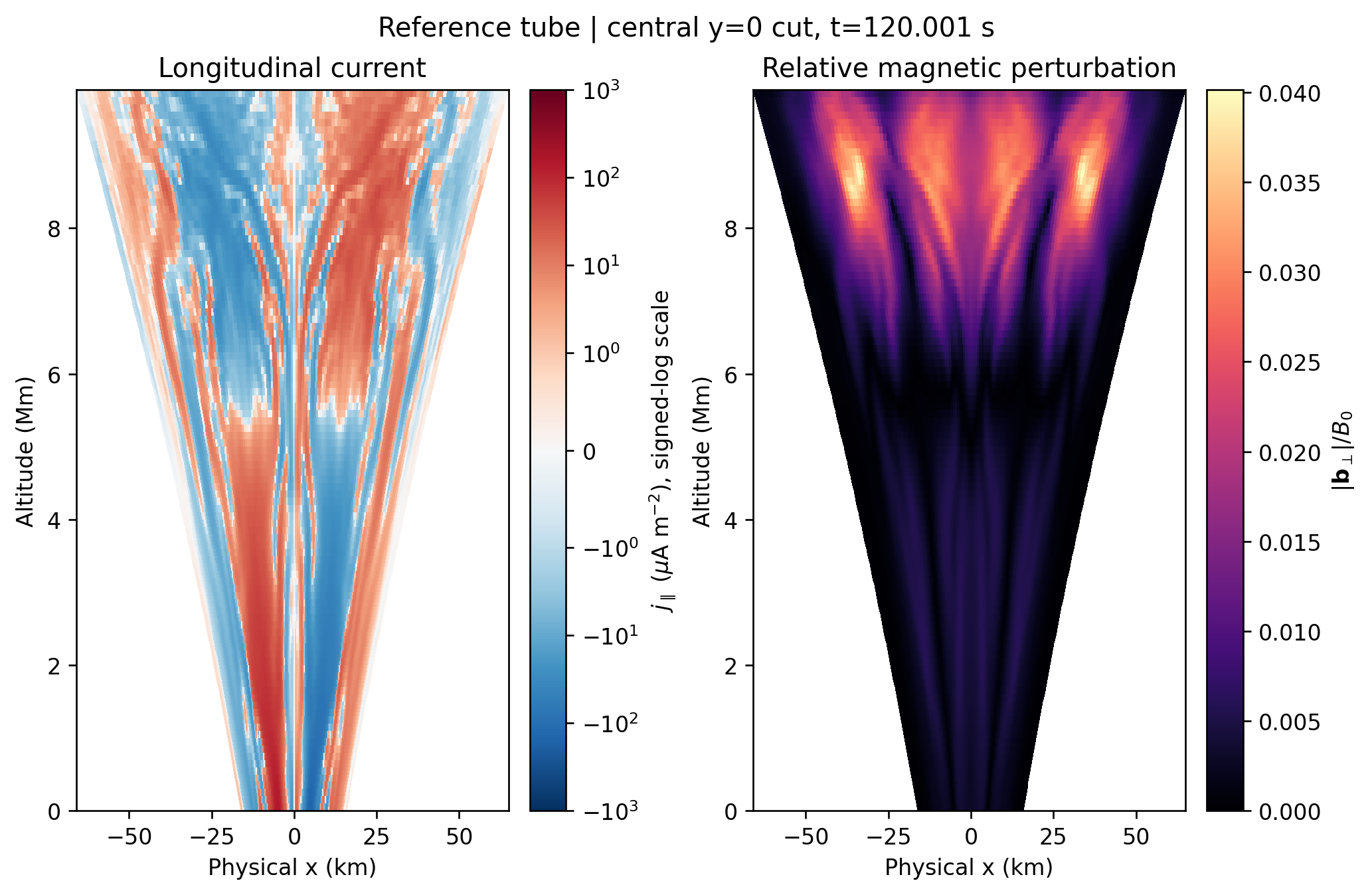}
\caption{Central $y=0$ longitudinal slice 
of signed current (left) and relative transverse magnetic perturbation
(right) at $t=120$~s.
The transverse dimension is exaggerated relative to the longitudinal dimension for visibility.}
\label{fig:longitudinal}
\end{figure}

The longitudinal section in Fig.~\ref{fig:longitudinal} places these
transverse structures within the full expanding tube. It compares
$j_\parallel$ with $|\mathbf b_\perp|/B_0$ in the central $y=0$ plane
at $120$~s. A large current indicates a strong transverse magnetic
gradient whereas a large $|\mathbf b_\perp|/B_0$ identifies a strong
perturbation relative to the local background.

The current concentrations also depend on numerical diffusion. Across
the uniform density runs, increasing the perpendicular diffusion
coefficients reduces the largest sampled current magnitudes 
by about $52\%$. At $50$~km altitude the
time-averaged spatial RMS current decreases from $25.4$ to
$19.8\ \mu\mathrm{A\,m^{-2}}$. Appendix~\ref{app:structure} gives
the time and diffusion comparisons. The narrowest features span only
a few grid spacings and so their physical widths are not established by
these images.

\FloatBarrier
\subsection{Potentials, fields and current gradients}
\begin{figure}[htbp]
\centering
\includegraphics[width=\textwidth,height=0.72\textheight,keepaspectratio]{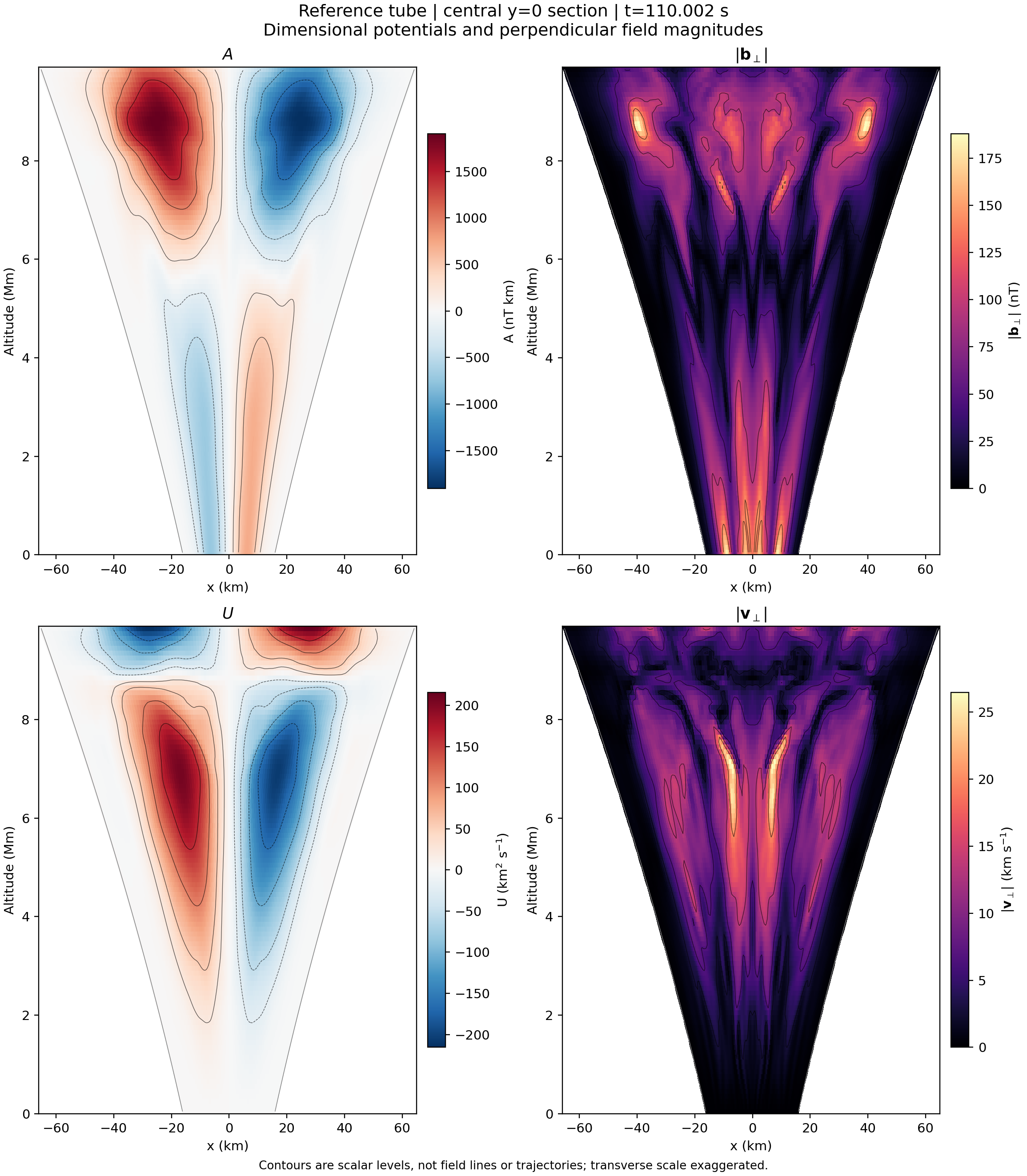}
\caption{Central $y=0$ longitudinal sections of the reference tube at
$t=110$~s: (top) $A$ and $|\mathbf b_\perp|$, (bottom) $U$ and
$|\mathbf v_\perp|$. Negative potential contours
are dashed. The velocity potential is interpolated between adjacent
longitudinal nodes to the magnetic field heights before evaluating the
perpendicular velocity. Perpendicular means of the potentials
are subtracted. The transverse
dimension is exaggerated relative to the longitudinal dimension.}
\label{fig:potentials-longitudinal}
\end{figure}

Figure~\ref{fig:potentials-longitudinal} compares the magnetic and velocity potentials
with the corresponding perpendicular field magnitudes in the central
$y=0$ plane at $110$~s. The potentials show broad, coherent structures
while their perpendicular derivatives reveal narrower spatial variations.
At each height, the Fourier amplitudes satisfy
$|\mathbf b_{\perp,\mathbf k}|=k_\perp|A_{\mathbf k}|$ and
$|\mathbf v_{\perp,\mathbf k}|=k_\perp|U_{\mathbf k}|$, while
$|j_{\parallel,\mathbf k}|=k_\perp^2|A_{\mathbf k}|/\mu_0$.
Taking derivatives thus gives increasing weight to small-scale components, i.e., 
broad potential contours can coexist with much finer field and current
structure. The longitudinal sections also show that the magnetic and
velocity structures have different altitude dependence.

At fixed height, $A$ is a transverse magnetic-flux function since
$\mathbf b_\perp\cdot\nabla_\perp A=0$. Its contours follow the transverse
magnetic perturbation, and differences in $A$ measure transverse magnetic
flux per unit length along the tube. Similarly, contours of $U$ follow the
instantaneous perpendicular flow.

\begin{figure}[htbp]
\centering
\includegraphics[width=\textwidth]{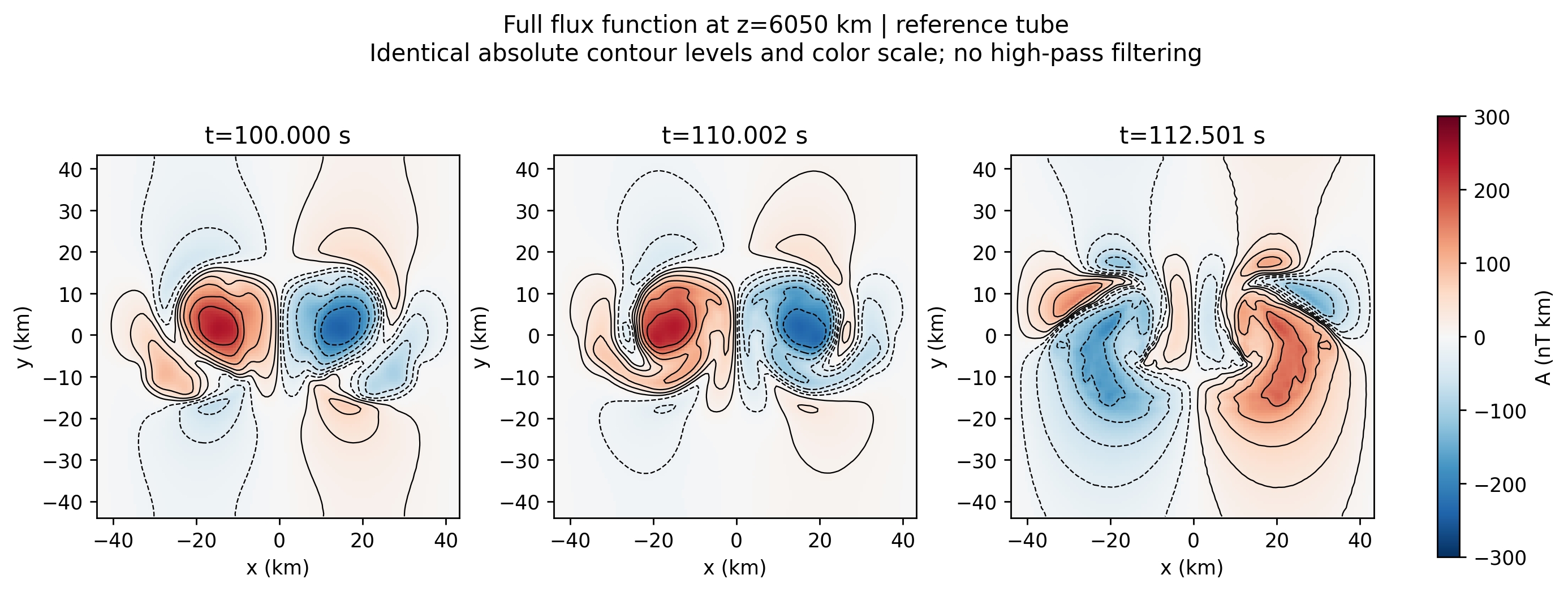}
\caption{Reference magnetic flux function at $6050$~km at
$100$, $110$ and $112.5$~s. Negative contours are dashed.
No high-pass filtering or spatial smoothing is applied.}
\label{fig:flux-structure}
\end{figure}

Figure~\ref{fig:flux-structure} shows $A$ at $6050$~km,
an altitude of strong nonlinear activity, at
three sampled times. A perpendicular mean is subtracted which changes only 
an additive offset in $A$.
The contours develop additional lobes and distortions showing that the
transverse magnetic pattern evolves beyond a fixed two-lobe shape.

\begin{figure}[htbp]
\centering
\includegraphics[width=\textwidth,height=0.72\textheight,keepaspectratio]{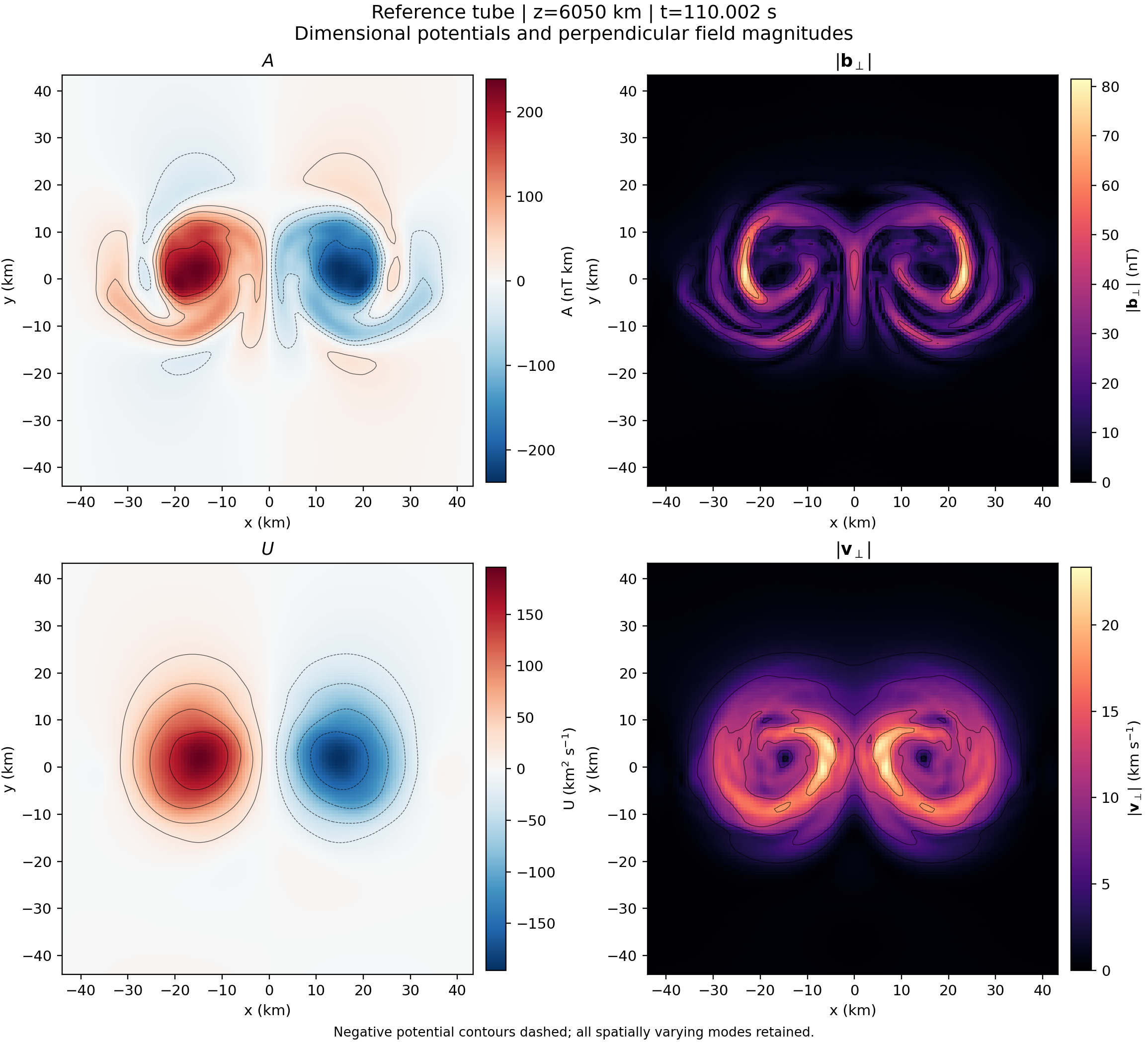}
\caption{Transverse sections of the reference tube at $6050$~km and
$t=110$~s: $A$ and $|\mathbf b_\perp|$ (top), and $U$ and
$|\mathbf v_\perp|$ (bottom).
Negative potential contours are dashed.
}
\label{fig:potentials-transverse}
\end{figure}

Figure~\ref{fig:potentials-transverse} compares the potentials and field
magnitudes at the same height at $110$~s. The field magnitude plots show
narrower bands than the potentials.

Across nine sampled times between $100$ and $120$~s, shells with index $m\ge7$ contain $1.52$--$5.57\%$ of the
$A$ variance but $25.4$--$50.3\%$ of the transverse magnetic
energy at this height. The variance is measured about the transverse-plane
mean of $A$. As a Fourier mode contributes in proportion to $|A_{\mathbf k}|^2$
to this variance but $k_\perp^2|A_{\mathbf k}|^2$ to magnetic energy, 
small scale structure can contribute significantly to magnetic
energy while remaining less prominent in the $A$ contours.

\FloatBarrier
\subsection{Structure with the non-uniform density profile}
\begin{figure}[htbp]
\centering
\includegraphics[width=.92\textwidth]{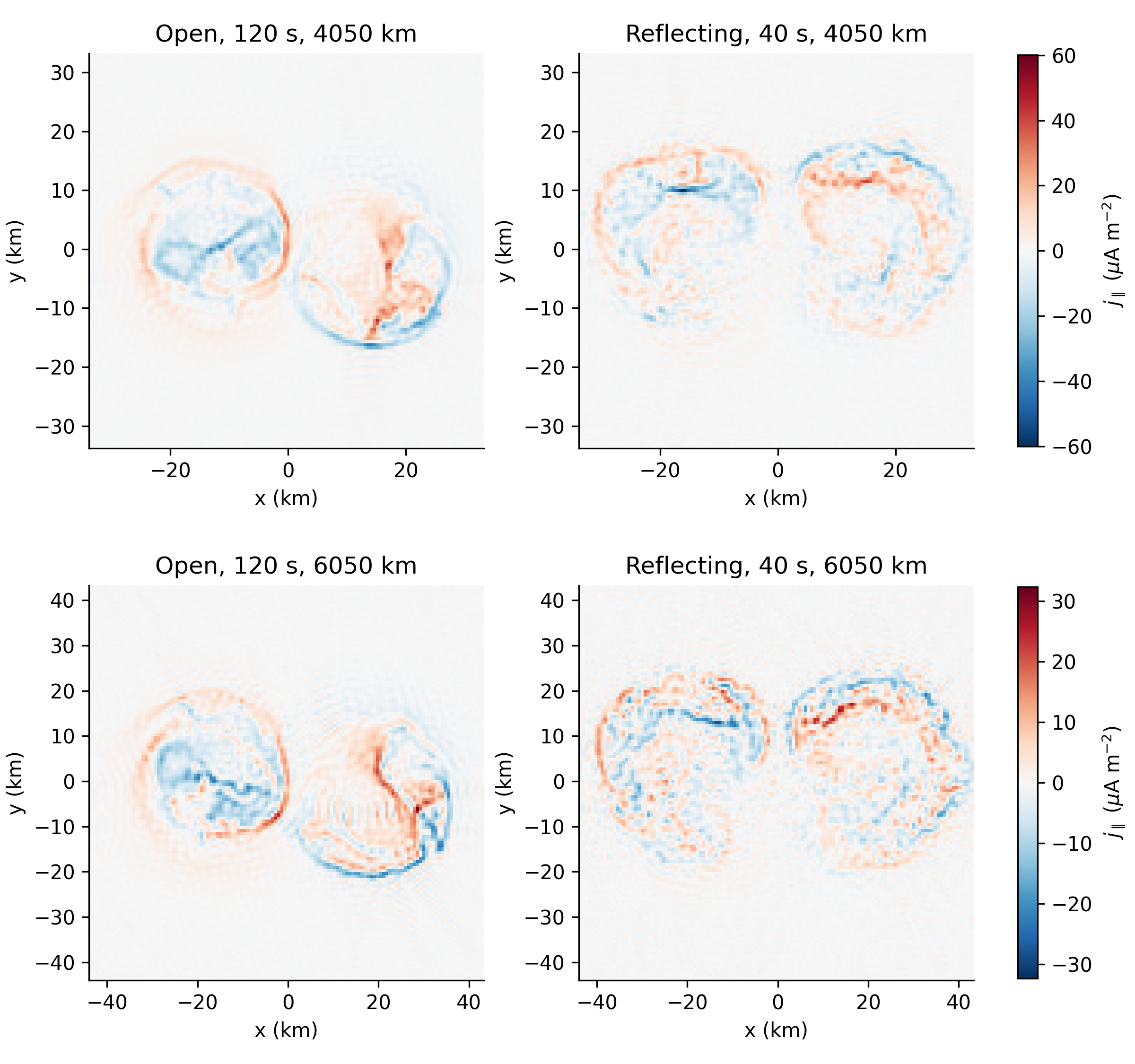}
\caption{Field-aligned current in the non-uniform density tube at $4050$~km
(top) and $6050$~km (bottom), for the open run at $120$~s (left) and the
reflecting run at $40$~s (right).
}
\label{fig:density-current}
\end{figure}

With the non-uniform density profile, both top boundary cases retain
the broad two-lobe organization of the driver while developing curved
current concentrations and smaller-scale variations around them.
Figure~\ref{fig:density-current} compares the transverse currents at
$4050$ and $6050$~km. The reflecting case at $40$~s shows more widespread
fine structure than the open case at $120$~s. The matched $30$--$40$-s energy budgets in
Section~\ref{sec:density-transfer} separately show greater dissipation
without greater mean stored energy in the reflecting run.

\begin{figure}[htbp]
\centering
\includegraphics[width=\textwidth]{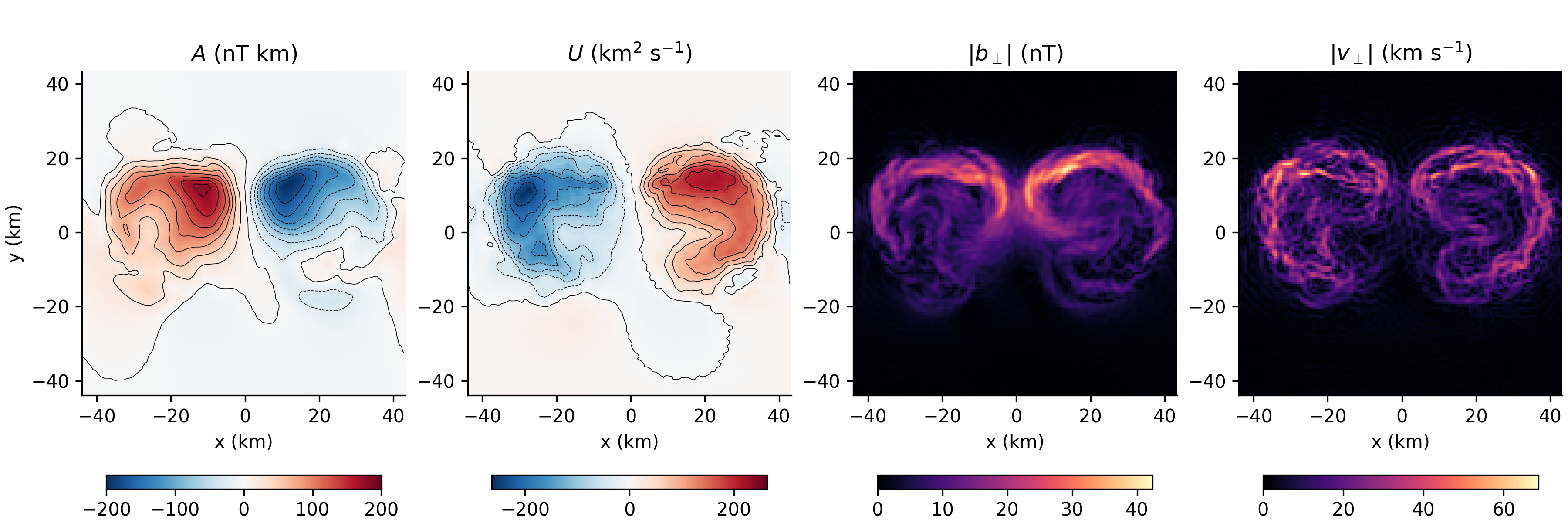}
\caption{Reflecting density profile run at $6050$~km and $40$~s:
magnetic potential $A$, velocity potential $U$ and their perpendicular
field magnitudes. Negative
contours are dashed.}
\label{fig:density-potentials-xy}
\end{figure}

The potential and field plots of the reflecting run in
Fig.~\ref{fig:density-potentials-xy} show the same distinction between
broad organization and fine structure. The two dominant $A$ and $U$
lobes retain the spatial pattern imposed by the bipolar driver in
Eq.~\eqref{eq:driver}, with distortions and secondary features within
and around them. The magnetic and velocity magnitudes contain narrower
variations as expected from the effect of the $k_\perp$ factors. 
Closed transverse $A$ contours follow circulating magnetic perturbations and
closed $U$ contours follow the instantaneous perpendicular flow.

\begin{figure}[htbp]
\centering
\includegraphics[width=\textwidth]{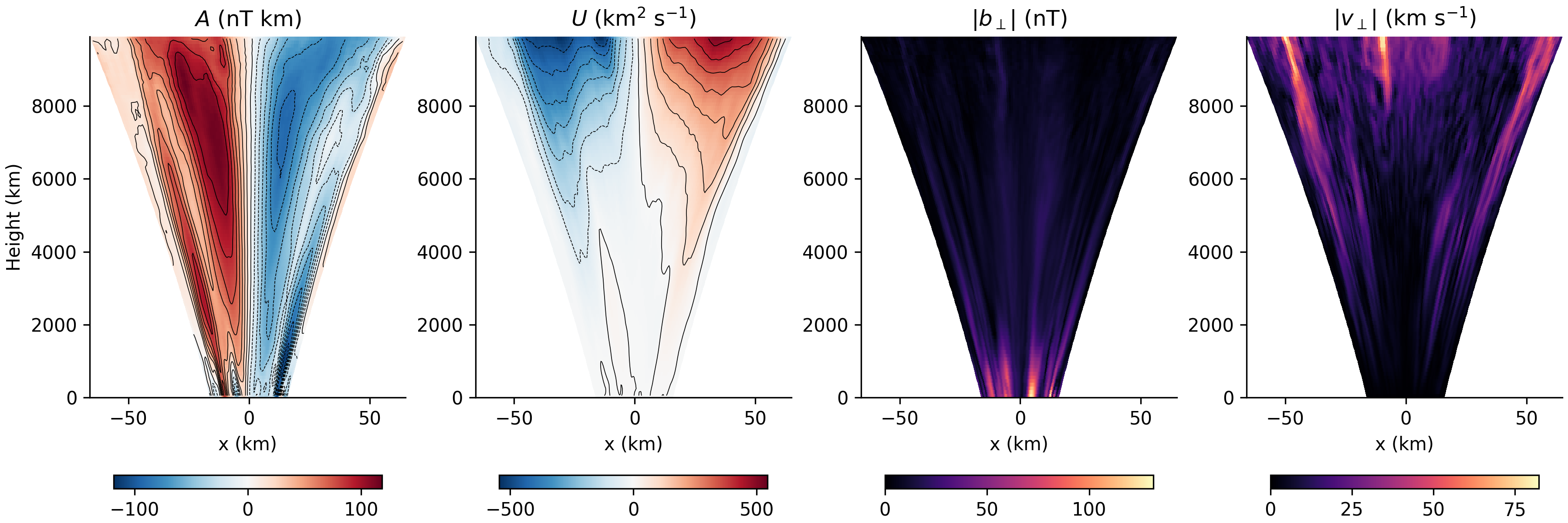}
\caption{Central $y=0$ longitudinal slice of the reflecting density profile
run at $40$~s with the same quantities as
Fig.~\ref{fig:density-potentials-xy}. The physical transverse mesh widens
with the tube and is exaggerated relative to the longitudinal scale.
$U$ is interpolated to the magnetic field heights.}
\label{fig:density-potentials-xz}
\end{figure}

The longitudinal slice in Fig.~\ref{fig:density-potentials-xz} shows how
this broad organization extends along the expanding tube. The field
magnitudes reveal narrower variations than the potentials.

\begin{figure}[htbp]
\centering
\includegraphics[width=.8\textwidth]{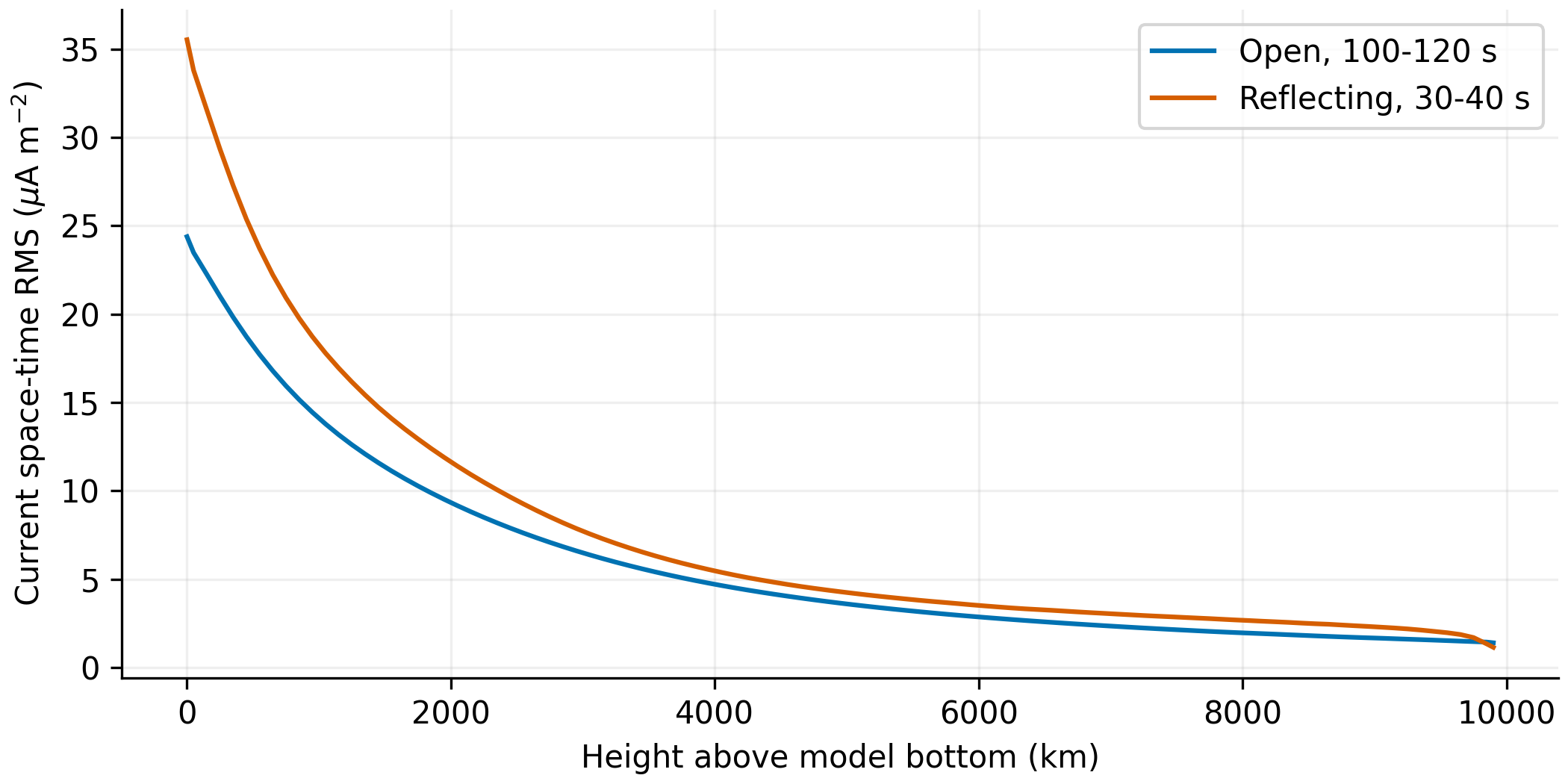}
\caption{Space--time RMS over transverse planes of field-aligned current in the non-uniform density
tube -- open top over $100$--$120$~s and reflecting top over $30$--$40$~s.}
\label{fig:density-current-rms}
\end{figure}

Figure~\ref{fig:density-current-rms} compares the current strength over
complete transverse planes rather than at selected peaks. The RMS is
$[\langle\langle j_\parallel^2\rangle_\perp\rangle_t]^{1/2}$ with the
spatial average taken over each complete transverse plane. The reflecting
run has higher RMS current over most of the tube in the plotted windows
with both profiles being strongest near the dense lower end. The averages
cover $30$--$40$~s for the reflecting case and $100$--$120$~s for the
open case.

\FloatBarrier
Together, these comparisons show that broad magnetic and flow patterns can
contain much finer current structure. The spectral transfer reported
earlier provides evidence of nonlinear redistribution across transverse
scales and the maps show the resulting spatial organization. They do
not identify tearing or reconnection and the narrowest
current widths remain sensitive to resolution and numerical diffusion.

\section{Summary and future work}
The main result of this study is the developing of altitude-dependent,
concentrated field-aligned currents in a driven, expanding magnetic flux
tube. The electron-inertial RMHD model follows these structures together
with the magnetic and velocity perturbations. Broad patterns in the
potentials coexist with finer structure in their perpendicular gradients,
most prominently in the current. The narrowest features are sensitive
to numerical resolution and diffusion, so their widths cannot yet be
assigned a physical dissipation scale.

Adding a density profile with a dense ionospheric end produces
a more realistic Alfv\'en speed profile for the auroral zone that first rises with height and then falls in the
expanding upper tube. In this stratified model, partial reflection at the
upper boundary produces finer current structure and stronger small-scale
activity than the open boundary.

The spectra and nonlinear transfer support this picture of current
structuring. Energy is transferred towards smaller perpendicular scales
with approximately power law intervals spanning up to $0.75$ decades in
the uniform density simulations and $0.83$ decades in the stratified
open boundary configuration. Their slopes and extent depend on height and
numerical diffusion. We do not find a broad constant flux inertial range
or establish a universal spectral index. The uniform wide slab comparison shows that 
more energy is retained at large perpendicular scales, but due to its different
background field and injected power, this contrast cannot be attributed 
to expansion alone.

Further work should determine how the current structures respond to
different density profiles, compositions, driving conditions  
and ionospheric conductances. Comparing nonlinear interaction
times with parallel propagation times and measuring parallel coherence
would test whether critical balance is an appropriate description of the
fluctuations \citep{Schekochihin2009}. A fully dipolar geometry would extend the present treatment
to include field-line curvature. Evolving density and parallel transport
would extend the prescribed-background model. Connecting the structured
currents to auroral electron precipitation requires an electron transport
or kinetic model.

\section*{Acknowledgements}
We thank Charles Seyler and John Dorelli for helpful discussions.
Computing resources were provided by the Minnesota Supercomputing Institute and 
our work was supported by NASA grant 80NSSC26K0281 to the University of Minnesota - Twin Cities.

\section*{Open Research}
The source code, run parameters, selected simulation outputs and analysis
scripts supporting this study are archived in Zenodo
\citep{MajmudarLysak2026Software}.
The software is distributed under the
MIT License and the data and documentation are under the
Creative Commons Attribution 4.0 International (CC BY 4.0) license.

\section*{Declaration of AI use}
Large language models such as OpenAI’s ChatGPT and Anthropic’s Claude were used to assist with 
code development and preparation of tables and figures.

\appendix
\section{Current morphology across time and numerical diffusion}
\label{app:structure}
We show some illustrative selections from simulation results of the current structure.

\begin{figure}[htbp]
\centering
\includegraphics[width=0.85\textwidth]{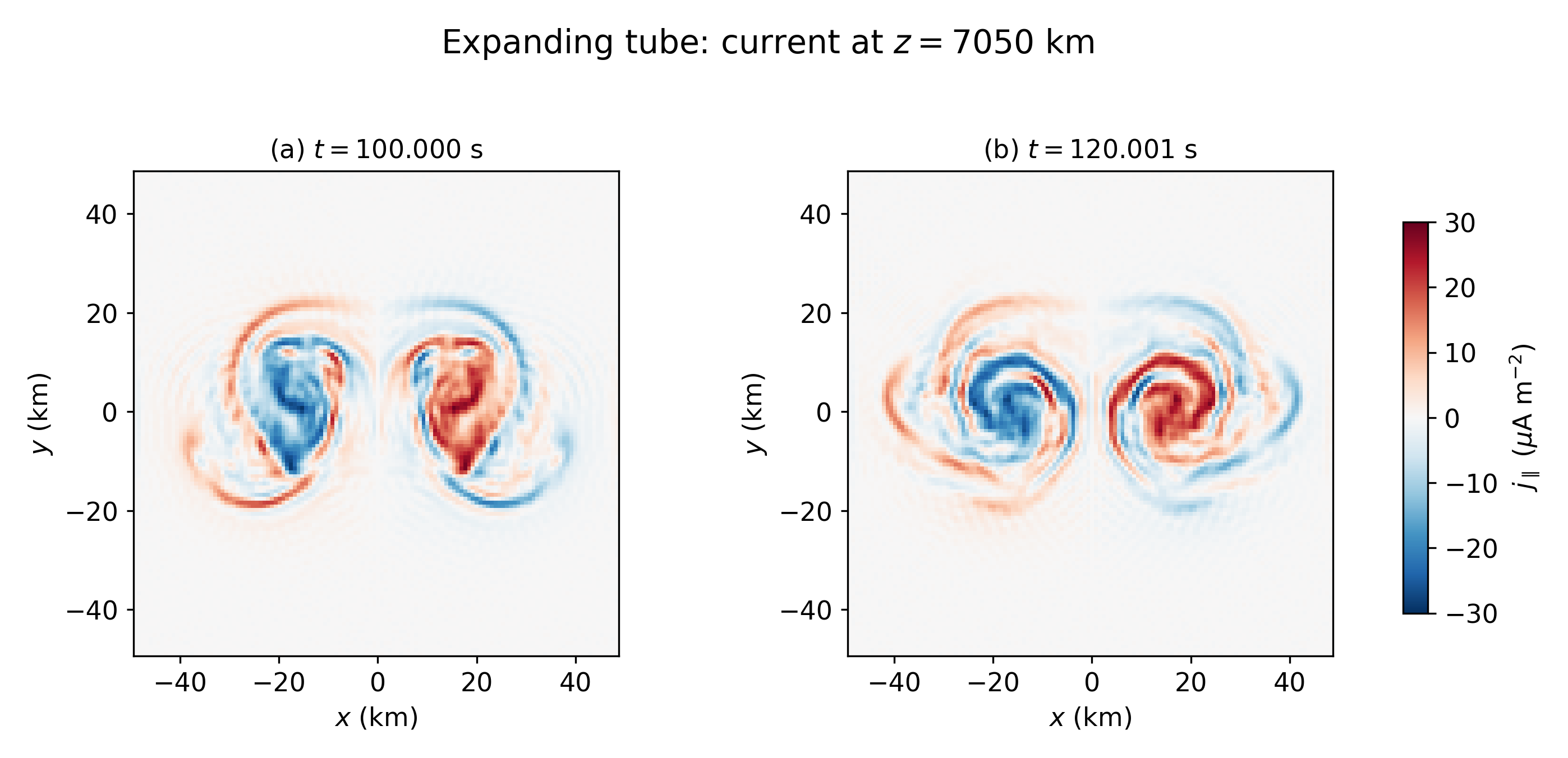}
\caption{Reference current at $7050$~km at $100.000227$ and
$120.001006$~s, approximately two forcing periods apart.
The estimated full widths at half maximum 
across the strongest current peak in each panel are $7.44$ and $4.90$ cells.}
\label{fig:current-time}
\end{figure}

\begin{figure}[htbp]
\centering
\includegraphics[width=\textwidth]{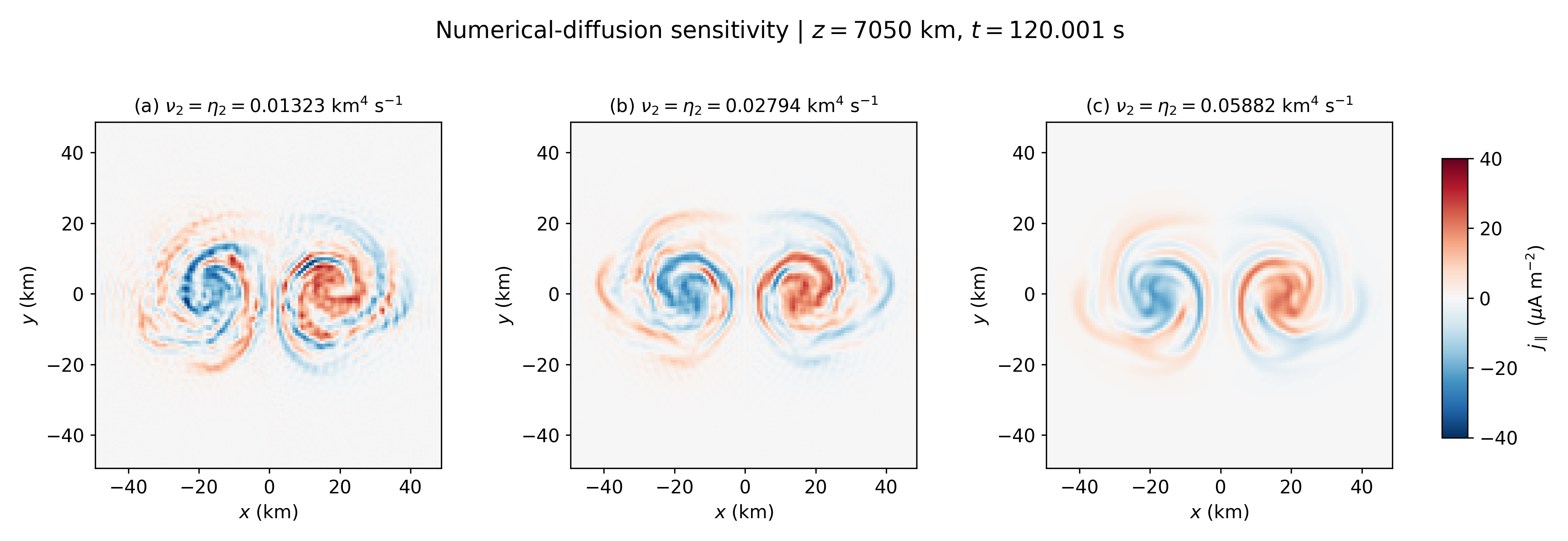}
\caption{Current at $7050$~km and $120$~s for
$\nu_2=\eta_2=0.0132$, $0.0279$ and $0.0588~\mathrm{km^4\,s^{-1}}$
(left to right).
Estimated full widths at half 
maximum across the strongest current peak in each panel are
$1.36$, $4.90$ and $8.79$ cells.}
\label{fig:current-diffusion}
\end{figure}
\FloatBarrier

\section{Selection of descriptive spectral intervals}
\label{app:band-selection}
For the uniform-density results in Table~\ref{tab:bands}, we examine all
contiguous shell intervals starting at $m\ge4$ containing at least eight
shells and spanning at least $0.5$ decades. The upper fitted shell is
restricted to $\lfloor0.8N_x/3\rfloor=34$ for $N_x=128$, leaving a margin
below the circular spectral cutoff. This fitting margin is only an analysis choice.
There are 87 candidate intervals at each height.
Each interval is fitted separately in
the $80$--$100$ and $100$--$120$~s averaged spectra.

We fit a straight line using least squares for $y_i=\log_{10}(E_i/E_*)$ versus
$x_i=\log_{10}m_i$, where $E_*$ is a fixed reference energy that affects
only the intercept. In each window, we require the fitted slope to be negative,
$R^2\ge0.98$, and the RMS log residual
$[N^{-1}\sum_i(y_i-\widehat y_i)^2]^{1/2}$ to not exceed $0.05$ decades,
where $N$ is the number of shells and $\widehat y_i$ is the fitted line.
To limit curvature we split the interval at the shell closest to its
midpoint and require the
two fitted slopes to differ by at most $0.3$. The full interval slopes
must differ by at most $0.2$ between time windows. We report the widest
passing interval at each height and ties are resolved by the smaller maximum
RMS log residual across the two windows.

\bibliographystyle{unsrtnat}
\bibliography{references}

\end{document}